\documentclass{aa}
\pdfoutput=1
\usepackage{natbib}
\usepackage{graphicx}
\usepackage{url}
\usepackage{txfonts}

\bibpunct{(}{)}{;}{a}{}{,} 

\usepackage{color}

\graphicspath{{./figs/}}

\begin{document}

\title{Long term variability of Cygnus X-1:} 
\subtitle{VI. Energy-resolved \mbox{X-ray} variability 1999--2011}

\titlerunning{Long term variability of Cygnus X-1: VI. Energy-resolved
  \mbox{X-ray} variability 1999--2011}

\author{V. Grinberg et al.}

\author{ \mbox{V.~Grinberg\inst{\ref{affil:remeis},\ref{affil:mit}}}
  \and \mbox{K.~Pottschmidt\inst{\ref{affil:cresst},\ref{affil:gsfc}}}
  \and \mbox{M.~B\"ock\inst{\ref{affil:mpifr}}} \and
  \mbox{C.~Schmid\inst{\ref{affil:remeis}}} \and
  \mbox{M.~A.~Nowak\inst{\ref{affil:mit}}} \and
  \mbox{P.~Uttley\inst{\ref{affil:uva_pannekoek}}} \and
  \mbox{J.~A.~Tomsick\inst{\ref{affil:ssl_berkeley}}} \and
  \mbox{J.~Rodriguez\inst{\ref{affil:aim_cnrs}}} \and
  \mbox{N.~Hell\inst{\ref{affil:remeis},\ref{affil:llnl}}} \and
  \mbox{A.~Markowitz\inst{\ref{affil:remeis},\ref{affil:ucsd},\ref{affil:humb}}}
  \and \mbox{A.~Bodaghee\inst{\ref{affil:ssl_berkeley}}}\and
  \mbox{M.~Cadolle Bel\inst{\ref{affil:universe}}}\and
  \mbox{R.~E.~Rothschild\inst{\ref{affil:ucsd}}} \and
 \mbox{J.~Wilms\inst{\ref{affil:remeis}} }
  }
\offprints{V.~Grinberg,\\ e-mail: {victoria.grinberg@fau.de}}
\institute{
Dr.\ Karl-Remeis-Sternwarte  and Erlangen Centre for
Astroparticle Physics (ECAP), Friedrich Alexander Universit\"at
Erlangen-N\"urnberg, Sternwartstr.~7, 96049 Bamberg, Germany
\label{affil:remeis}
\and
Massachusetts Institute of Technology, Kavli Institute for Astrophysics, Cambridge, MA 02139, USA
\label{affil:mit}
\and
CRESST, University of Maryland Baltimore County, 1000 Hilltop Circle,
Baltimore, MD 21250, USA \label{affil:cresst}
\and
NASA Goddard Space Flight Center, Astrophysics Science Division, Code
661, Greenbelt, MD 20771, USA \label{affil:gsfc}
\and
Max-Planck-Institut f\"ur Radioastronomie, Auf dem H\"ugel 69, 53121
Bonn, Germany \label{affil:mpifr}
\and
Astronomical Institute ``Anton Pannekoek'', University of Amsterdam,
Kruislaan 403, Amsterdam, 1098 SJ, The Netherlands
\label{affil:uva_pannekoek}
\and Space Sciences Laboratory, 7 Gauss Way, University of California,
Berkeley, CA 94720-7450, USA \label{affil:ssl_berkeley} \and
Laboratoire AIM, UMR 7158, CEA/DSM - CNRS - Universit\'{e} Paris
Diderot, IRFU/SAp, 91191 Gif-sur-Yvette, France \label{affil:aim_cnrs}
\and Lawrence Livermore National Laboratory, 7000 East Ave.,
Livermore, CA 94550, USA \label{affil:llnl} 
\and 
Center for
Astrophysics and Space Sciences, University of California San Diego,
La Jolla, 9500 Gilman Drive, CA 92093-0424, USA \label{affil:ucsd}
\and 
Alexander von Humboldt fellow \label{affil:humb}
\and Ludwig-Maximilians University, Excellence Cluster ``Universe'',
Boltzmannstr. 2 85748 Garching, Germany \label{affil:universe}
}
\date {Received: --- / Accepted: ---}

\abstract{We present the most extensive analysis of Fourier-based
  \mbox{X-ray} timing properties of the black hole binary Cygnus X-1 to date,
  based on 12 years of bi-weekly monitoring with RXTE from 1999 to
  2011. Our aim is a comprehensive study of timing behavior across all
  spectral states, including the elusive transitions and extreme hard
  and soft states. We discuss the dependence of the timing properties
  on spectral shape and photon energy, and study correlations between
  Fourier-frequency dependent coherence and time lags with features in
  the power spectra. Our main results are: (a) The fractional rms in
  the 0.125--256\,Hz range in different spectral states shows complex
  behavior that depends on the energy range considered. It reaches its
  maximum not in the hard state, but in the soft state in the
  Comptonized tail above 10\,keV. (b) The shape of power spectra in
  hard and intermediate states and the normalization in the soft state
  are strongly energy dependent in the 2.1--15\,keV range. This
  emphasizes the need for an energy-dependent treatment of power
  spectra and a careful consideration of energy- and mass-scaling when
  comparing the variability of different source types, e.g., black
  hole binaries and AGN. PSDs during extremely hard and extremely soft
  states can be easily confused for energies above $\sim$5\,keV in the
  0.125--256\,Hz range. (c) The coherence between energy bands drops
  during transitions from the intermediate into the soft state but
  recovers in the soft state. (d) The time lag spectra in soft
  and intermediate states show distinct features at frequencies
  related to the frequencies of the main variability components seen
  in the power spectra and show the same shift to higher frequencies
  as the source softens. Our results constitute a template for other
  sources and for physical models for the origin of the \mbox{X-ray}
  variability. In particular, we discuss how the timing properties of
  \mbox{Cyg~X-1} can be used to assess the evolution of variability with
  spectral shape in other black hole binaries. 
 Our results suggest that none of the available theoretical models
  can explain the full complexity of \mbox{X-ray} timing behavior of \mbox{Cyg~X-1},
  although several ansatzes with different physical assumptions are
  promising.}


\maketitle

\section{Introduction}

The canonical states of accreting black hole binaries have first been
observed and defined in the spectral domain: a hard state with a power
law spectrum with a photon spectral index of $\sim$1.7 and a soft
state with a spectrum dominated by thermal emission from an accretion
disk. They are joined by usually short-lived transitional or
intermediate states.  The whole sequence of states -- from hard state
over the intermediate into soft and then again into intermediate and
finally into hard state -- can best be depicted on a
hardness-intensity-diagram (HID), where transient sources that undergo
a full outburst follow a \textsf{q}-shaped track \citep[][see
\citeauthor*{McClintock_Remillard_2006a},
\citeyear{McClintock_Remillard_2006a}, for a different
nomenclature]{Fender_2004a}. Radio emission is detected in the hard
state, with jets imaged for \object{\mbox{Cyg~X-1}}
\citep{Stirling_2001a} and \object{GRS~1915$+$105}
\citep{Dhawan_2000a,Fuchs_2003a}. In the soft state, radio emission is
strongly quenched. Evidence for similar spectral behavior also exists
in several other classes of accreting objects such as neutron star
\mbox{X-ray} binaries \citep[e.g.,][\object{Aql~X-1}]{Maitra_2004a},
active galactic nuclei \citep[AGN;][]{Koerding_2006a}, and dwarf novae
\citep[][\object{SS~Cyg}]{Koerding_2008a}.

The spectral states, including the different flavors of the
intermediate state, show distinct \mbox{X-ray} timing characteristics, such
as shapes of power spectra or time lags between emission at different
energies.  Timing parameters seem to be a remarkably sensitive tool
for defining state transitions
\citep[e.g.,][]{Pottschmidt_2003b,Fender_2009a,Belloni_2010a}.

While the radio emission, and possibly also the gamma-ray emission
above 400\,keV \citep{Laurent_2011a,Jourdain_2012a} originate in the
jet, the origin of the \mbox{X-ray}s is still unclear. As shown, e.g., by
\citet{Nowak_2011a}, the combination of the best resolution and the
most broadband \mbox{X-ray} spectra available today fails to enable us to
statistically distinguish between jet models \citep{Markoff_2005a,
  Maitra_2009a} and thermal and/or hybrid Comptonization in a corona
\citep[e.g.,][]{Coppi_1999a,Coppi_2004a}. The fluorescent Fe K$\alpha$
line and a reflection hump point towards a contribution by
reflection, independent of the origin of the continuum
\citep[see][for a review and \citealt{Duro_2011a},
\citealt{Tomsick_2014a}, and references therein for
\mbox{Cyg~X-1}]{Reynolds_2003a}.

Spectro-timing analysis holds the promise of solving this ambiguity,
as a truly physical model has to consistently describe both spectral
and timing behavior. While no self-consistent models that would
encompass all parameters exist yet, some studies address, e.g.,
simultaneous modeling of photon spectra and root mean square
variability (rms) spectra \citep{Gierlinski_2005a} or photon spectra
and Fourier-dependent time lags \citep{Cassatella_2012a}. Theoretical
ansatzes to describe the \mbox{X-ray} variability include propagating mass
accretion rate fluctuations \citep[usually based on][see, e.g.,
\citealt{Ingram_2013a} for an analytical model]{Lyubarskii_1997a},
up-scattering in a jet \citep{Reig_2003a,Kylafis_2008a}, and full
magnetohydrodynamic simulations \citep[][who concentrate on spectra
but also address timing properties]{Schnittman_2013a}.

There is mounting evidence for similarities in timing behavior between
\mbox{X-ray} binaries, AGN \citep[see][for a review]{McHardy_2010a},
and recently also cataclysmic variables \citep[see][for discovery of
Fourier-dependent time lags]{Scaringi_2013a}. Because of their
brightness and short variability time scales, \mbox{X-ray} binaries
remain the best laboratories to investigate these phenomena and are
key to deciphering the complex and currently highly disputed interplay
of accretion and ejection processes. We note especially that AGN are
usually viewed on a single state due to the very much longer variation
timescale, and thus \mbox{X-ray} binaries are needed to study the
variety of states and their inter-relationships.

The first step to understanding the \mbox{X-ray} timing
characteristics is a fundamental overview of their evolution with the
spectral shape that can only be achieved with a high number of high
quality observations densely covering all states, including the
elusive transitions. Most previous works concentrate on
energy-independent evolution of rms and power spectral distributions
(PSDs) with spectral state
\citep[e.g.,][]{Pottschmidt_2003b,Belloni_2005a,Axelsson_2006a,Klein-Wolt_2008a},
although further spectral shape-, energy-, and Fourier
frequency-dependent correlations have been noted in individual
observations and smaller samples
\citep[e.g.,][]{Homan_2001a,Rodriguez_2002b,Kalemci_2004a,Rodriguez_2004a,
  Boeck_2011a,Cassatella_2012b,Stiele_2013a}, often with a focus on
the behavior of narrow quasi-periodic oscillations.  Missing are
consistent analyses of the energy-resolved evolution of rms and PSDs,
Fourier-frequency dependent evolution of cross-spectral quantities
(coherence function and lags), and correlations of features in PSDs
and Fourier-frequency dependent cross-spectral quantities over the
full range of spectral states and over multiple transitions. In this
paper, we address these questions with an extraordinarily long and
well sampled RXTE campaign on the high mass black hole binary
\mbox{Cyg~X-1} that enables us to conduct the most comprehensive
spectro-timing analysis of a black hole binary to date.

Located at a distance of $1.86^{+0.12}_{-0.11}$\,kpc
\citep[][consistent with \citealt{Xiang_2011a}]{Reid_2011a}, \mbox{Cyg~X-1}
is bright (in the hard state
$\sim$$7\times10^{-9}$\,erg\,cm$^{-2}$\,s$^{-1}$
in the 1.5--12\,keV band of RXTE-ASM) and persistent and therefore a
prime target for both spectral and timing studies. Often considered a
prototypical black hole binary, it is frequently used for comparisons
with other black hole binaries \citep[e.g,][]{Munoz-Darias_2010a} and
AGN \citep{Markowitz_2003a,McHardy_2004a,Papadakis_2009a}.  Although
the spectrum of the source is never fully dominated by the disk and
the bolometric luminosity only changes by a factor of $\sim$4
\citep[][and references
therein]{Cui_1997a,Shaposhnikov_2006a,Wilms_2006a}, \mbox{Cyg~X-1}
often undergoes state transitions
\citep[e.g.,][]{Pottschmidt_2003b,Wilms_2006a,Grinberg_2013a} and is
therefore also well suited for studies of the intermediate states.

Here, we analyze data from the full set of RXTE observations of
\mbox{Cyg~X-1}. This paper is a part of a series where we previously
analyzed spectro-timing correlation in the hard state 1998 to 2001
\citep{Pottschmidt_2003b}, the rms-flux relation
\citep{Gleissner_2004a}, the radio-\mbox{X-ray} correlations
\citep{Gleissner_2004b}, the spectral evolution 1999--2004
\citep{Wilms_2006a}, and states and state transitions 1996--2012 with
all sky monitors \citep{Grinberg_2013a}. We start
Sect.~\ref{sect:data} by introducing the data and the general behavior
of the source during the time period covered by this analysis and
follow with a description of the spectral analysis and employed
\mbox{X-ray} timing techniques. In Sect.~\ref{sect:rms_and_power}, we
discuss the energy independent and dependent evolution of the rms and
the PSDs with spectral shape. In Sect.~\ref{sect:cross}, we discuss
the evolution of cross-spectral quantities with spectral shape. We
address the implication of our results for the analysis of other
sources in Sect.~\ref{sect:other} and for theories explaining the
origin of \mbox{X-ray} variability in
Sect.~\ref{sect:theory}. Sect.~\ref{sect:summary} summarizes our
finding with a focus on the implied directions for further
investigations.

\section{Data Analysis}\label{sect:data}

The data analyzed here are mostly from a bi-weekly observational
campaign with RXTE that was initiated by some of us in 1999 and
continued until the demise of RXTE at the end of 2011. Parts of the
data have been analyzed in the previous papers of this series
\citep{Pottschmidt_2003b,Gleissner_2004a,Gleissner_2004b,Wilms_2006a,Grinberg_2013a}
and by other authors
\citep[e.g.,][]{Axelsson_2005a,Axelsson_2006a,Shaposhnikov_2006a}. For
all pointed RXTE observations of \mbox{Cyg~X-1} made during the
lifetime of the RXTE satellite (MJD 50071--55931), we extracted
spectral data in the \texttt{standard2f} mode from Proportional
Counter Unit 2 of RXTE's Proportional Counter Array
\citep[PCA,][]{Jahoda_2006a} and from RXTE's High Energy \mbox{X-ray}
Timing Experiment \citep[HEXTE,][]{Rothschild_1998a} on a satellite
orbit by satellite orbit basis. The data reduction was performed with
HEASOFT 6.11 as described by \citet{Grinberg_2013a} and resulted in a
total of 2741 spectra. At the time of writing there had been no
changes to relevant pieces of HEASOFT since this release. We stress
the importance of the orbit-wise approach, as spectral and timing
properties of \mbox{Cyg~X-1} can change on time scales of less than a
few hours \citep{Axelsson_2005a,Boeck_2011a,Grinberg_2013a}.

\subsection{Long-term source behavior}

\begin{figure*}
\includegraphics[width=\textwidth]{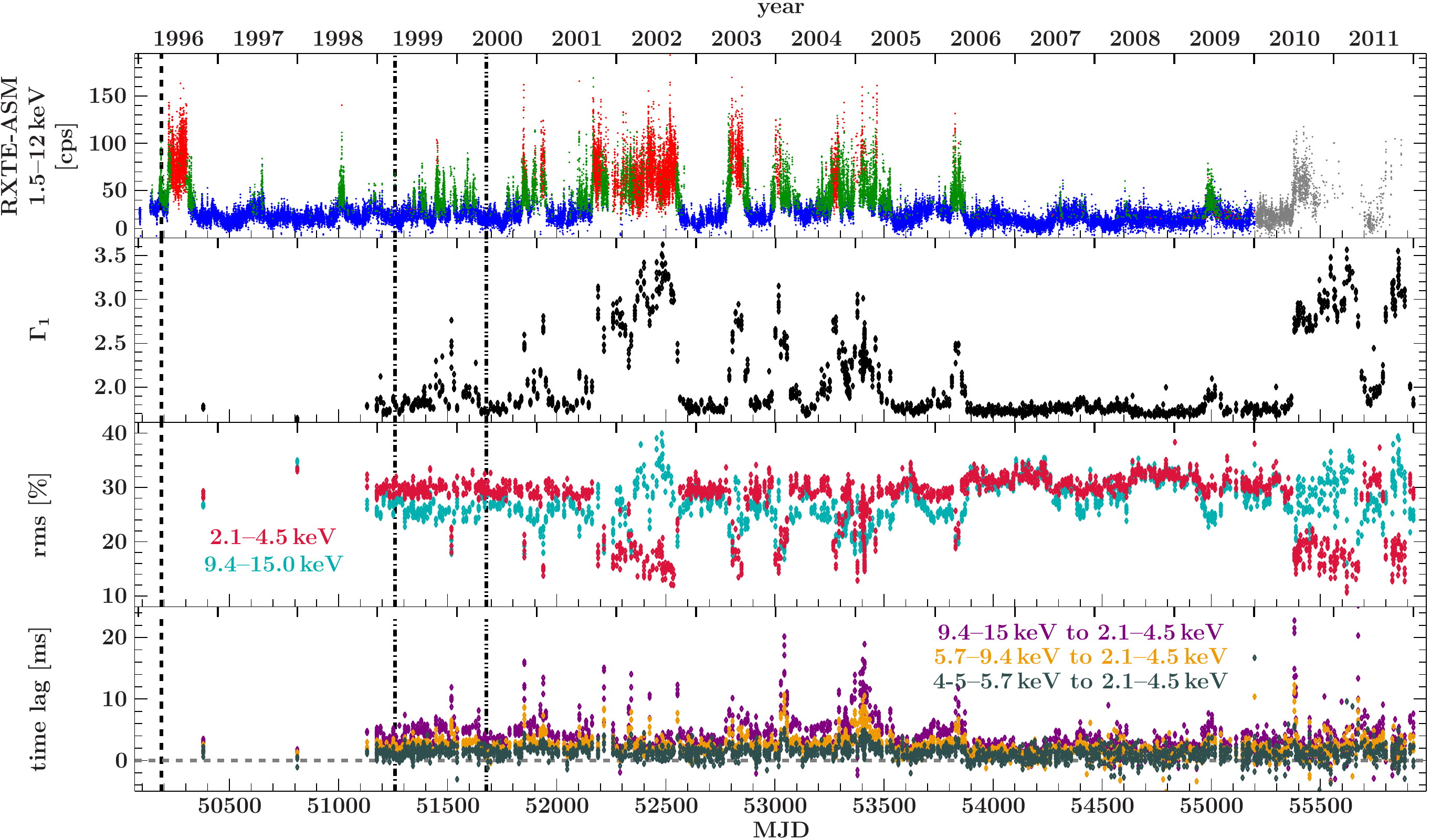}
\caption{Evolution of \mbox{Cyg~X-1} over the RXTE lifetime. Vertical
  lines represent the starting times of the RXTE calibration epochs
  used in this work: dashed line epoch 3, dot-dashed line epoch 4 and
  dot-dot-dashed line epoch 5. The total ASM count rate is color-coded
  according to ASM-based state definition of \citet{Grinberg_2013a},
  which uses both, ASM count rate and ASM hardness, of a given
  measurement: blue represents the hard state, green the intermediate
  state, and red the soft state. ASM data points after MJD 55200 are
  affected by instrumental decline
  \citep{Vrtilek_2013a,Grinberg_2013a} and shown in gray. The soft
  photon index $\Gamma_1$ is shown only for those 1980 RXTE
  observations that were conducted in the
  \texttt{B\_2ms\_8B\_0\_35\_Q} binned data mode, see
  Sect.~\ref{sect:timing}. The rms is calculated in the 0.125--256\,Hz
  range (Sect.~\ref{sect:rms}), the time lags are averaged values in
  the 3.2--10\,Hz range
  (Sect.~\ref{sect:avglags}).}\label{fig:overview}
\end{figure*}

The general behavior of \mbox{Cyg~X-1} during the lifetime of RXTE has
been discussed by \citet{Grinberg_2013a}. Figure~\ref{fig:overview}
presents an overview of the evolution of the RXTE All Sky Monitor
\citep[ASM,][]{Levine_1996a} count rate, the spectral shape and a
choice of typically used \mbox{X-ray} timing parameters.

The data used here cover periods of different source behavior such as
pronounced, long hard and soft states, and multiple failed and full
state transitions. The strict use of the same data mode means that we
do not cover some of the observations included in previous long-term
monitoring analyses by \citet{Pottschmidt_2003b},
\citet{Axelsson_2005a, Axelsson_2006a}, or \citet{Shaposhnikov_2006a},
especially the data from the extreme hard state before 1998 May. The
large time span covered in this work that includes the extraordinarily
long, hard state of $\sim$MJD~53900--55375 (mid-2006 to mid-2010) and
the following series of stable soft states represents, however, a
major improvement over all previous analyses.

\subsection{Spectral analysis}\label{sect:spec}

The spectral analysis presented here is the same that we used in
\citet[][Sects.~2.2 and 3.2]{Grinberg_2013a}, so we give only a brief
overview. We describe the $\sim$2.8--50\,keV PCA data and the
18--250\,keV HEXTE data, both rebinned to a signal to noise ratio of
10, using a simple phenomenological model that has been shown to offer
the best description of RXTE data of \mbox{Cyg~X-1} \citep[][who also
discuss advantages of this model over more physically motivated
approaches]{Wilms_2006a}. The basic continuum model consists of a
broken power law with soft photon index $\Gamma_1$, hard photon index
$\Gamma_2$, and a break energy at $\sim$10\,keV. The continuum is
modified by a high energy cut-off ($E_{\mathrm{cut}} \approx
20$--30\,keV, $E_{\mathrm{fold}} \approx 100$--300\,keV, both
correlated with $\Gamma_1$), an iron K$\alpha$-line at about 6.4\,keV
and absorption, described by the
\texttt{tbnew}\footnote{\url{http://pulsar.sternwarte.uni-erlangen.de/wilms/research/tbabs/}}
model, an advanced version of \texttt{tbabs}, with abundances of
\citet{Wilms_2000a} and the cross sections of \citet{Verner_1996a}. An
additional soft excess is modeled by a multicolor disk
\citep[\texttt{diskbb},][]{Mitsuda_1984a,Makishima_1986} where
necessary, i.e., predominantly in softer observations. A good fit can
be achieved with at least one of the two models for all our
observations. The disk is accepted as real if the improvement in
$\chi^2$ is more than 5\%, irrespective of the
$\chi^2_\mathrm{red}$-value of the model without the disk. This
approach ensures a smooth transition between models without and with a
disk. Note that the dominance of systematic errors in the lowest
channels which define the disk component prevents us from using a
significance based criterion. PCA's calibration uncertainties are
taken into account by adding systematic errors in quadrature to the
data (1\% added to the fourth PCA bin and 0.5\% to the fifth PCA bin
in epochs 5 and 4; 1\% to the fifth PCA bin and 0.5\% to the sixth PCA
bin in epoch 3, see \citealt{Hanke_2011_PhD} and
\citealt{Boeck_2011a}).

We identify seven observations where the above $\chi^2$ criterion for
the model selection results in a preference for the model with a disk
but the disk is peculiarly strong and the soft photon index is very
steep, making these observations outliers in the tight correlation of
$\Gamma_1$ with $\Delta \Gamma = \Gamma_1 - \Gamma_2$
\citep{Wilms_2006a}. The observations can also be clearly seen as
outliers in the tight correlation between rms-ratio and $\Gamma_1$
discussed in Sect.~\ref{sect:rms} when $\Gamma_1$-values of the models
with a disk are used. When modeling these seven observations without a
disk all have pronounced residuals at $\sim$5\,keV due to the Xe
L-edge \citep{Wilms_2006a}, no signature of a strong disk in the
residuals, and an acceptable reduced $\chi^2 < 1.3$. This behavior
indicates that the strong disk component and the steeper soft photon
index compensate the instrumental Xe L-edge and are not an adequate
description of the source spectrum. We therefore accept the model
without a disk as the best fit model for these
observations\footnote{In our previous study \citep{Grinberg_2013a},
  these seven observations did not receive any special treatment. This
  does, however, not influence any of our earlier conclusions.}. The
$\Gamma_1$-values for these observations shift from the 2.0--2.2 to
the 1.8--2.0 range. We emphasize that if the disk were real and not
compensating for sharp calibration features, the removal of the disk
component would result in an increase of $\Gamma_1$, contrary to what
is seen here.

\begin{figure}
\centering
\resizebox{\hsize}{!}{\includegraphics{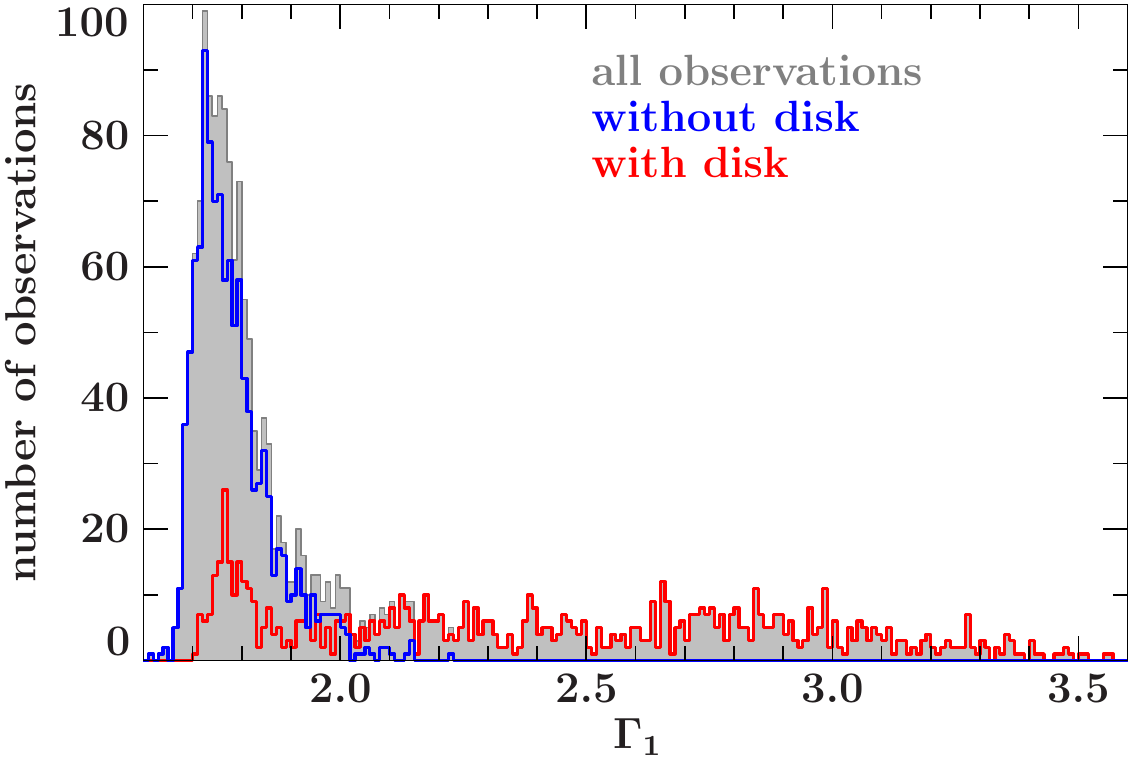}}
\caption{Number of orbit-wise RXTE observations of \mbox{Cyg~X-1} in the
  \texttt{B\_2ms\_8B\_0\_35\_Q} mode with a given $\Gamma_1$. Shown
  are observations that require a disk (red), that do not require a
  disk (blue), and the total number of observations
  (gray).}\label{fig:gamma}
\end{figure}

We consider the 1980 spectra for which data in the
\texttt{B\_2ms\_8B\_0\_35\_Q} mode is available (see
Sect.~\ref{sect:timing}). Our fits yield $\chi^2_\mathrm{red}<2.5$ for
all observations and a good fit with $\chi_\mathrm{red}^2<1.2$ can be
achieved for 1721 out of these 1980.  In the following, we will use
$\Gamma_1$ as a proxy for the spectral state, since all other spectral
parameters show strong correlations with $\Gamma_1$
\citep{Wilms_2006a}. In particular, in \citet{Grinberg_2013a} we have
defined the hard state as $\Gamma_1 < 2.0$, the intermediate state as
$2.0 < \Gamma_1 < 2.5$, and the soft state as $\Gamma_1 >
2.5$. Figure~\ref{fig:gamma} shows the number of observations at a
given $\Gamma_1$; note the good coverage of the soft and intermediate
states, but also the very high number of observations in the hard
state.

\subsection{Calculation of \mbox{X-ray} timing quantities}\label{sect:timing}

We extracted light curves which are strictly simultaneous with the
spectral data. For a consistent treatment of the timing properties, we
only used observations in the \texttt{B\_2ms\_8B\_0\_35\_Q} mode,
i.e., binned data with 8 bands (channels 0--10, 11--13, 14--16,
17--19, 20--22, 23--26, 27--30, and 31--35) and with an intrinsic time
resolution of $2^{-9}\,\mathrm{s}\approx\,2\,\mathrm{ms}$, resulting
in 1980 RXTE orbits. Because this data mode does not include PCU
information, light curves can only be extracted from all PCU units
that were operating during a particular observation together. Since
dead time effects depend on the number of active detectors, we treat
each combination of PCUs separately. This approach results in an
increase of the number of light curves to 2015, some of them
associated with the same spectrum. All analyses in this paper were
performed with ISIS~1.6.2
\citep{Houck_Denicola_2000a,Houck_2002,Noble_Nowak_2008a}.

\begin{table*}
  \caption{Energy bands used for different RXTE calibration  epochs.}\label{tab:bands}
\centering
\begin{tabular}{lcccc}
\hline
\hline
epoch & band 1 & band 2 & band 3 & band 4\\
&energy [keV] (channels)&energy [keV] (channels)&energy [keV] (channels)&energy [keV] (channels)\\
\hline
3& $\sim$1.9\tablefootmark{b}--4.1 (0--10)& $\sim$4.1--5.1 (11--13)& 
$\sim$5.1--8.3 (14--22)& $\sim$8.3--13.0 (23--35)\\
4& $\sim$2.1\tablefootmark{b}--4.6 (0--10)& $\sim$4.6--5.9 (11--13)&$\sim$5.9--9.7 (14--22)& $\sim$9.7--15.2 (23--35)\\
5\tablefootmark{a} (PCU 0)& $\sim$2.0\tablefootmark{b}--4.6 (0--10)& $\sim$4.6--5.8 (11--13)&$\sim$5.8--9.7 (14--22)& $\sim$9.7--15.4 (23--35)\\
5\tablefootmark{a} (PCU 1,2,3,4)& $\sim$2.1\tablefootmark{b}--4.5 (0--10)& $\sim$4.5--5.7 (11--13)&$\sim$5.7--9.4 (14--22)& $\sim$9.4--14.8 (23--35)\\
\hline
\end{tabular}
\tablefoot{Values from
  \url{http://heasarc.nasa.gov/docs/xte/e-c_table.html}. Starting
  times of epochs are shown in Fig.~\ref{fig:overview}. 
  Since the majority of the data is from epoch 5, we cite the 
  band energies 
  throughout this paper as 2.1--4.5\,keV for band 1, 4.5--5.7\,keV for 
  band 2, 5.7--9.4\,keV 
  for band 3, and 9.4--15\,keV for band 4.
  \tablefoottext{a}{Epoch 5 is
    defined by loss of the propane layer in PCU 0 due to a
    micrometeorite impact.} \tablefoottext{b}{Nominally the
    0th channel extends to $\sim$0 keV, but the effective area is
    negligible below 2\,keV \citep{Jahoda_2006a}.}}
\end{table*}

In Table~\ref{tab:bands}, we list the channel and energy ranges for
the four energy bands that were used throughout this work for the
calculation of the timing properties (band~3 and band~4 combine three
channel bands each in order to obtain higher count rates). Most of our
data fall into epochs~4 and~5 (Fig.~\ref{fig:overview}). The energy
range covered by the bands is slightly time dependent because of
changes in the high voltage of the PCA, but such small changes will
not influence the qualitative analysis presented in this work:
\begin{itemize}
\item Band~1 (2.1--4.5\,keV) covers the range with a significant
  contribution from the accretion disk, which can dominate this band
  in the soft states. At the high soft count rates of the soft states
  this band can suffer from telemetry overflow
  \citep{Gleissner_2004a,Gierlinski_2008a}, resulting in artifacts at
  high frequencies above $\sim$30--50\,keV as can be seen, e.g., on
  Fig.~\ref{fig:avg_psds2} for $\Gamma_1 \sim 2.61$ and $\sim
  2.71$\footnote{Telemetry overflow is only a problem for the
    observations with the highest count rate, i.e., the observations
    at the intermediate to soft state transition
    \citep{Grinberg_2013a}. Not all observations at a given $\Gamma_1$
    are, however, affected since spectra with the same $\Gamma_1$ can
    vary by a factor of $\sim$4. For example for $2.5 < \Gamma_1 <
    2.7$, the absorbed flux in the 3--10\,keV range varies between 4
    and 14\,keV\,s$^{-1}$\,cm$^{-2}$.}.
\item Band~2 (4.5--5.7\,keV) covers the soft part of the spectrum
  above the main contribution of the disk. The band is below the
  prominent Fe K$\alpha$-line at $\sim$6.4\,keV, but since the line is
  broad its contribution may be significant
  \citep{Wilms_2006a,Shaposhnikov_2006a,Duro_2011a}.
\item Band~3 (5.7--9.4\,keV) includes the Fe K$\alpha$-line and covers
  the spectrum up to the spectral break at $\sim$10\,keV.
\item Band~4 (9.4--15\,keV) mainly covers the spectrum above the
  spectral break at $\sim$10\,keV and below the high-energy cut-off.
\end{itemize}

We calculated all timing properties (power spectra [power spectral
densities, PSD], cross power spectra, coherence and time lags)
following \citet{Nowak_1999a} and \citet{Pottschmidt_2003b}. For the
timing analysis each light curve was split into segments of
$n_\mathrm{bins} = 4096$\,bins, i.e., 8\,s length. Timing properties
were calculated for each segment and then averaged over all segments
of a given light curve using the appropriate statistics. The mean
number of segments used was $\sim$225 and the mean light curve length
thus $\sim$30\,min. Before rebinning, the values of all timing
properties were calculated for discrete Fourier frequencies $f_i$
linearly spaced every $\Delta f = 1/(n_\mathrm{bins} \Delta t) =
0.125\,\mathrm{Hz}$. The Nyquist frequency of the data is
$f_\mathrm{max} = 1/(2\Delta t) = 256\,\mathrm{Hz}$ and the lowest
frequency accessible is $f_\mathrm{min} = 1/(n_\mathrm{bins} \Delta t)
= 0.125\,\mathrm{Hz}$, i.e., we cover slightly over three decades in
temporal frequency. As shown by \citet{Nowak_1999a}, for a source such
as \mbox{Cyg~X-1} and using the approach described above, the dead
time corrected, noise-subtracted PSDs are reliable to at least
100\,Hz. The systematic uncertainties for the coherence function are
negligible below 30\,Hz and the phase and time lags are detectable in
the 0.1--30\,Hz range \citep[][but see also
Sect.~\ref{sect:cof}]{Nowak_1999a}.

Longer individual segments would enable us to probe lower Fourier
frequencies. The lower number of averaged segments per observation
would mean higher uncertainties, but generally still reliable results
in the case of PSDs. Coherence function and lags are, however, higher
order derivative statistics and therefore more sensitive to
uncertainties. Thus, to constrain them well, a larger number of
segments is necessary, i.e., the individual segments have to be
shorter.

For PSDs, we choose the normalization of \citet{Belloni_1990a} and
\citet{Miyamoto_1991a}, where the PSD is given in units of the squared
fractional rms variability per frequency interval. Since the variance
that each frequency contributes is given in units of average signal
count rate, this normalization is most suitable for comparing PSD
shapes independently of source brightness
\citep{Pottschmidt_2002_diss}. In order to illustrate the contribution
of the variability at a given Fourier frequency, $f_i$, to the total
variability, we show the PSDs in units of PSD times frequency,
$\mathrm{PSD}(f_i) \times f_i$ \citep{Belloni_1997a,Nowak_1999a}. We
calculated the fractional rms by summing up the contributions at
individual frequency ranges without employing any modeling of the PSD
shape; all rms values in this work are given for the 0.125--256\,Hz
range.

While the PSD described the variability of one light curve, the
relationship between two simultaneous light curves can be characterized
by the cross spectral density. Such a cross spectrum is calculated as
a product of the Fourier transform on one light curve with the complex
conjugate of the Fourier transform of the other light curve.

The coherence function, $\gamma^2 (f_i)$, a derivative of the norm of
the cross spectrum, measures the Fourier-frequency dependent degree of
linear correlation between two time series, in this case the light
curves \citep{Vaughan_1997a}. An intuitive geometrical interpretation
of $\gamma^2 (f_i)$ as a length of a vector sum in the complex plane
can be found in \citet{Nowak_1999a}.

For two correlated time series, we can define the Fourier-frequency
dependent phase lag $\phi_i$ at a frequency $f_i$ as the difference
between the phases of the Fourier transforms of the light curves in
both energy bands, calculated as the argument of the complex cross
spectrum. The time lag at the frequency $f_i$ is then given by
$\delta_t(f_i) = \phi_i/(2\pi f_i)$. Our sign convention is such that
the hard light curve lags the soft for a positive lag. Since the phase
lag, $\phi$, is defined on the interval $[-\pi,\pi[$, there is an
upper limit for absolute value of the time lag. Experience shows that
the lags are below this limit in the considered frequency range in
objects such as \mbox{Cyg~X-1} \citep[e.g.,][]{Nowak_1999a,
  Pottschmidt_2000a, Pottschmidt_2003b, Boeck_2011a}, \object{Swift
  J1753.5$-$0127} \citep{Cassatella_2012b}, or \object{XTE
  J1752$-$223} \citep{Munoz-Darias_2010a}.

\section{Rms and power spectra}\label{sect:rms_and_power}

\subsection{Evolution of fractional rms with spectral shape}\label{sect:rms}

We start with the most basic quantity, the rms
variability of the source. As we have previously shown
\citep{Grinberg_2013a}, the total fractional rms in the 0.125--256\,Hz
frequency band strongly depends on the spectral state and on the
energy band. Here, we analyze this behavior in more detail.

\subsubsection{Energy independent vs. energy dependent evolution of rms}

\begin{figure}
\centering
\resizebox{\hsize}{!}{\includegraphics{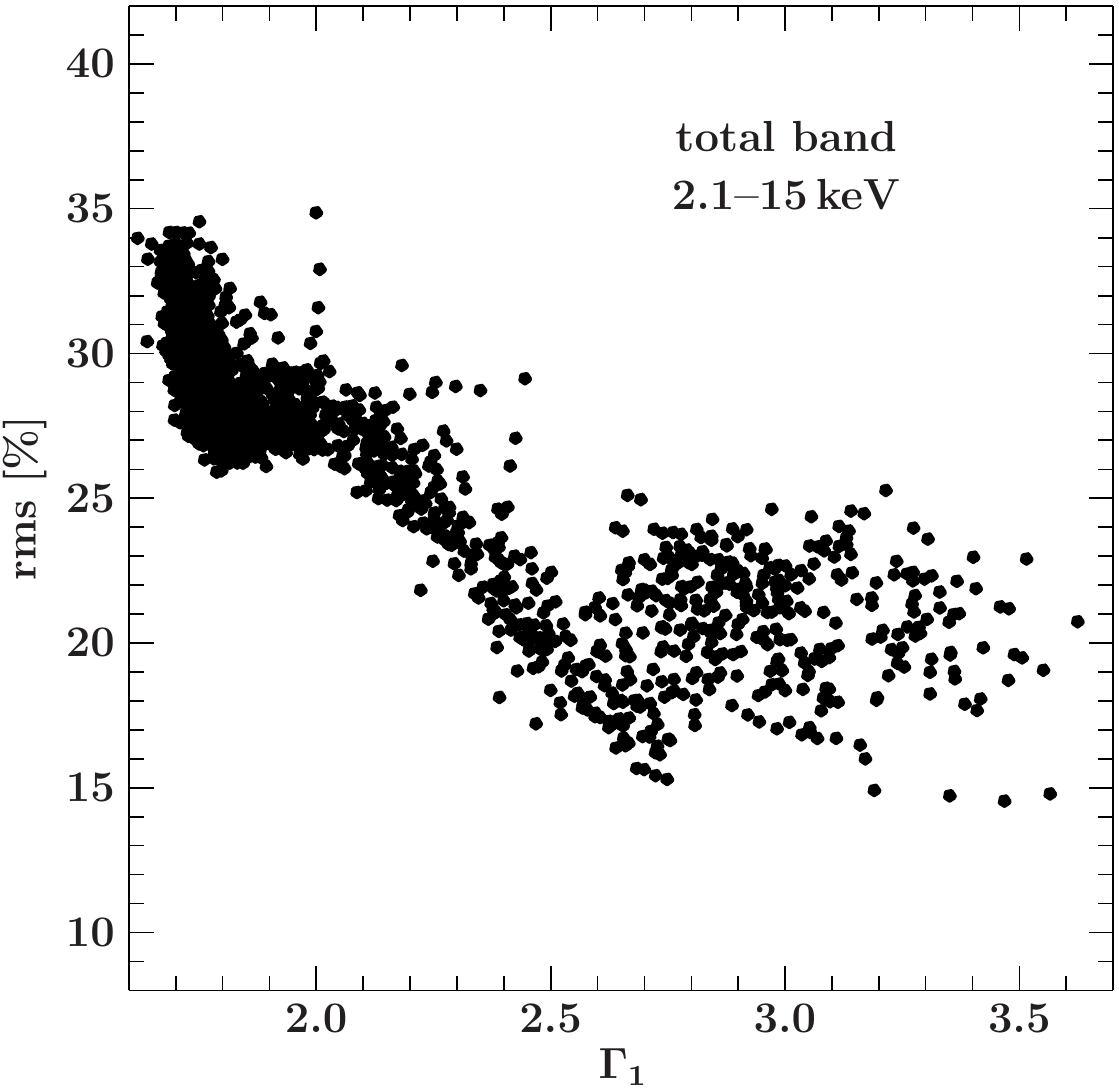}}
\caption{Fractional rms in the 2.1--15\,keV band vs. soft photon index
  $\Gamma_1$.}\label{fig:totalrms}
\end{figure}

Figure~\ref{fig:totalrms} shows how the fractional rms in the
2.1--15\,keV band depends on the spectral shape.  Here and in the
following Sections, the errors of the rms are expected to be smaller
than the spread of the shown correlations at any give $\Gamma_1$.  The
rms reaches its largest values of $\sim$34\% at the lowest $\Gamma_1$
of $\sim$1.6.  The $\Gamma_1$-rms correlation is negative below
$\Gamma_1 \approx 1.8$, then flat until $\Gamma_1 \approx 2.0$, again
negative until $\Gamma_1 \approx 2.5$--2.7, and finally flattens out
at an rms of $\sim$20\% but with a larger scatter.

\begin{figure*}
\includegraphics[width=\textwidth]{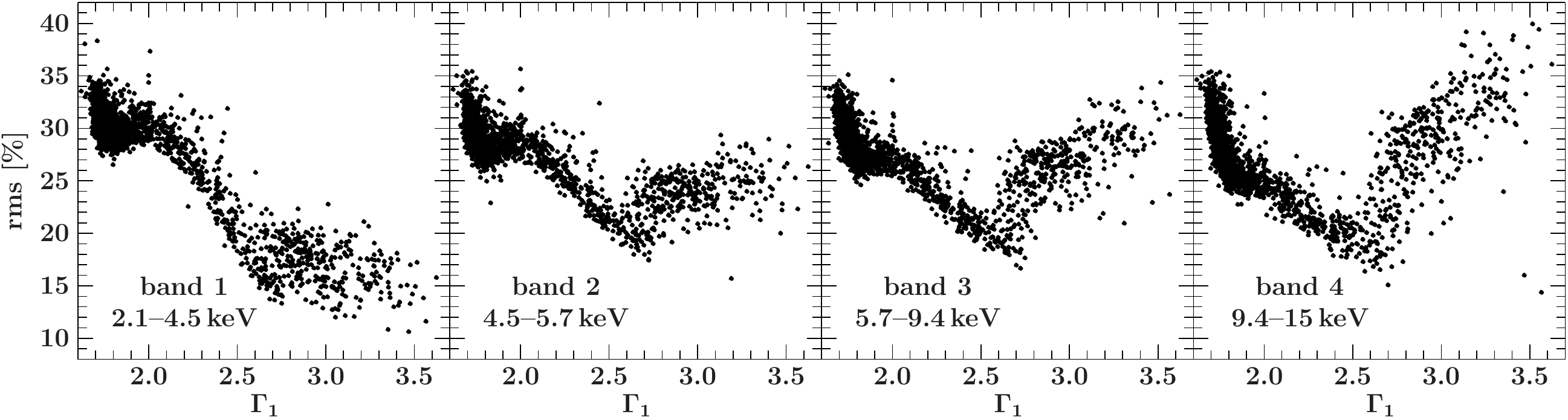}
\caption{Fractional rms in the 0.125--256\,Hz frequency band for
  different energy bands vs. the soft photon index, $\Gamma_1$, of
  the broken power law fits.}\label{fig:gamma_rms}
\end{figure*}

\begin{table}
\caption{Spearmann $\rho$ correlation coefficients for $\Gamma_1$-rms correlation.}\label{tab:rmsg1}
\centering
\begin{tabular}{ccccc}
\hline \hline
band &$\rho_{\Gamma_1 < 1.8}$&$\rho_{1.8 < \Gamma_1< 2.0}$&$\rho_{2.0 < \Gamma_1 < 2.65}$&
$\rho_{2.65 < \Gamma_1}$\\
\hline
total&$-0.61$&0.21\tablefootmark{b}&$-0.89$&0.02\tablefootmark{c}\\
1&$-0.50$&0.25&$-0.89$&$-0.28$\\
2&$-0.55$&0.19&$-0.88$&0.43\\
3&$-0.62$&$-0.064$\tablefootmark{a}&$-0.84$&0.65\\
4&$-0.64$&$-0.31$&$-0.70$&0.68\\
\hline
\end{tabular}
\tablefoot{All null hypothesis probabilities $P <10^{-4}$ except 
  for \tablefoottext{a}{with $P = 0.18$}, 
\tablefoottext{b}{with $P = 0.21$}, and \tablefoottext{c}{with $P = 0.66$}} 
\end{table}

The energy dependent fractional rms reveals where most of this
variability in the total 2.1--15\,keV band comes from in the different
spectral states. Figure~\ref{fig:overview} shows the temporal
evolution of the fractional rms in the soft band~1 and the hard band~4
(2.1--4.5\,keV and 9.4--15\,keV) over the lifetime of RXTE. During
hard states, the variability is high at about $\sim$30\% in both
bands. In the soft state, the rms drops to 10--20\% in band~1, but in
band~4 it is slightly larger than in the hard state. The intermediate
state is associated with a decrease of rms in both bands. This
behavior is better visualized in Fig.~\ref{fig:gamma_rms}, where we
plot rms as a function of $\Gamma_1$ for bands 1--4. By eye, we
identify four $\Gamma_1$ ranges with different behavior (see
Table~\ref{tab:rmsg1} for correlation coefficients for bands 1--4 and
the total band).
\begin{itemize}
\item $\Gamma_1 < 1.8$: the $\Gamma_1$-rms correlation is negative and
  becomes stronger at higher energies in the individual and the total
  bands;
\item $1.8 \le \Gamma_1< 2.0$: the correlation is positive for band~1
  and~2 and negative for band~4. In band~3, the rms values are
  consistent with being uncorrelated with $\Gamma_1$, an
  understandable behavior given the change of sign in adjacent
  bands. In the total band, there is no correlation;
\item $2.0 \le \Gamma_1 < 2.65$: the correlation is strong and
  negative in all bands including the total, though it becomes less
  steep with increasing energy;
\item $2.65 \le \Gamma_1$: the correlation changes from negative in
  band~1 (note the negative $\rho$-value in
    Table~\ref{tab:rmsg1}) to positive in bands~2--4 and becomes
  stronger with increasing energy. The total rms in the 2.1--15\,keV
  band shows no correlation with spectral shape. There is a slight
  indication of further structure at $\Gamma_1 \sim 2.8$--$2.9$ in
  energy bands~2--4.
\end{itemize}
The changes in slope of the rms-$\Gamma_1$ relationship at $\Gamma_1
\sim 1.8$ and $\Gamma_1 \sim 2$ are present both when considering only
observations that do not require a disk and only observations that do
require a disk. They are therefore not artifacts of the choice of
spectral model used to measure $\Gamma_1$. For $\Gamma_1 > 2$, the
number of observations which do not require a disk is negligible
(Fig.~\ref{fig:gamma}). Note also that the $\Gamma_1$ thresholds
represent approximate values, since the transition from one behavior
pattern to the other is not clearly defined.

\begin{figure}
  \resizebox{\hsize}{!}{\includegraphics{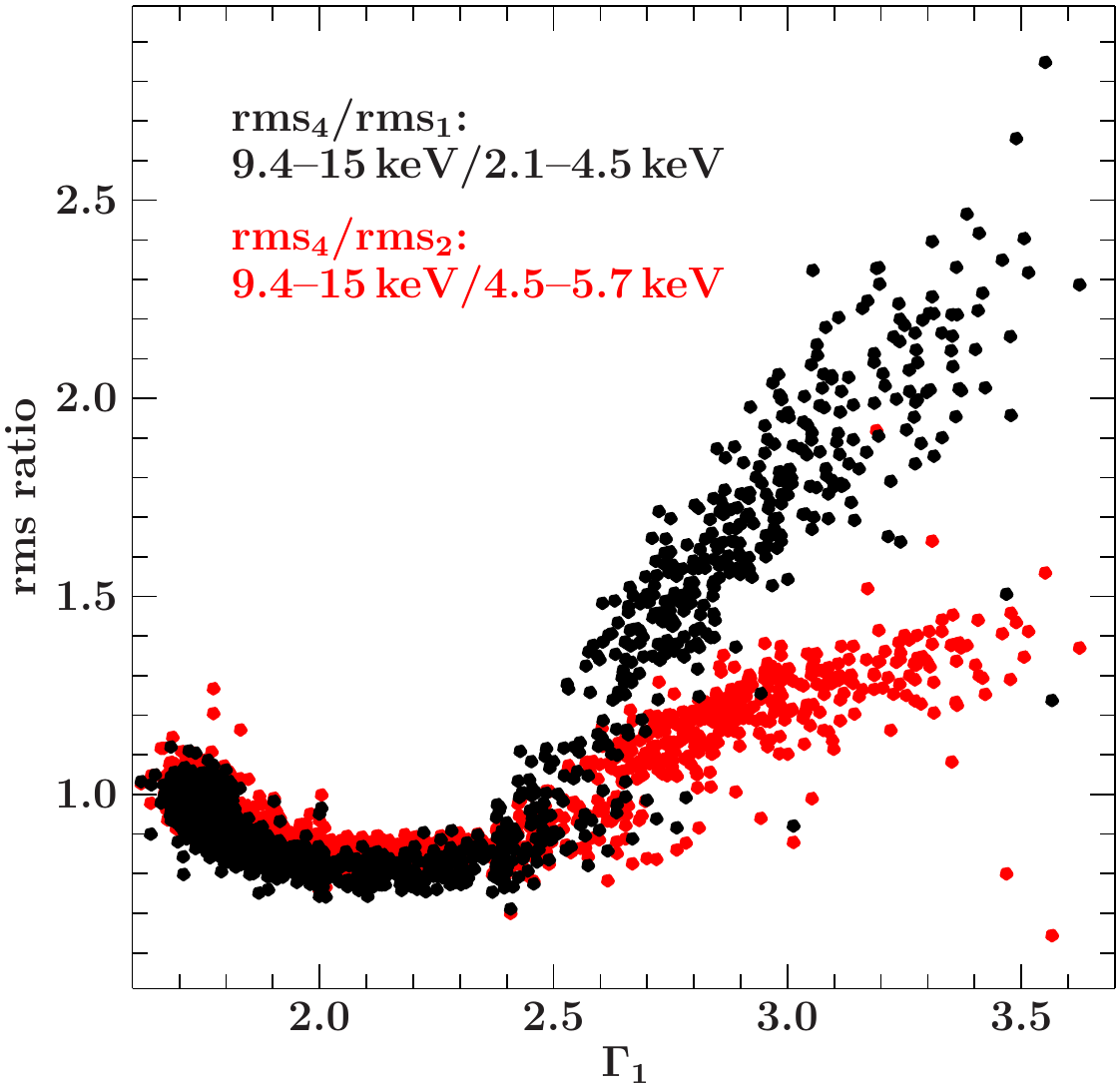}}
  \caption{Ratio of fractional rms in different energy bands (band~1
    -- $\mathrm{rms}_1$, band~2 -- $\mathrm{rms}_2$, band~4 --
    $\mathrm{rms}_4$,) vs. soft photon index
    $\Gamma_1$.}\label{fig:deltarms}
\end{figure}

A compact representation of the rms behavior is given by the ratio of
the rms values between the different bands
(Fig.~\ref{fig:deltarms}). The relation is tight enough that it can
serve as a simple check for the spectral shape of the source without
any model assumptions. We used it, e.g., to check our broken power law
fits for outliers as described above in Sect.~\ref{sect:spec}. We
emphasize that because of the strong energy dependence of the rms, the
total fractional rms in the 2.1--15\,keV band
(Fig.~\ref{fig:totalrms}) cannot be used to determine whether one soft
state observation is softer than an other in terms of relative
$\Gamma_1$, while the rms at energies above $\sim$6\,keV or the ratio
of rms at different energies (Fig.~\ref{fig:deltarms}) can.

The energy-independent analyses of RXTE data by \citet[][2.1--15\,keV
range]{Pottschmidt_2003b} and \citet[][2--9\,keV
range]{Axelsson_2006a} are consistent with the rms evolution presented
here, although different time ranges are covered. A decrease of the
rms as the source softens was also observed at higher energies, namely
in the 10--200\,keV range using \textsl{Suzaku}-PIN and
\textsl{Suzaku}-GSO by \citet{Torii_2011a} and in the 27--49\,keV band
using INTEGRAL-SPI by \citet{Cabanac_2011a}, although neither of them
observed a full soft state with $\Gamma_1 \gtrsim 2.5$, where our
analysis shows an increase of the rms above $\sim$6\,keV.

\begin{figure}
  \resizebox{\hsize}{!}{\includegraphics{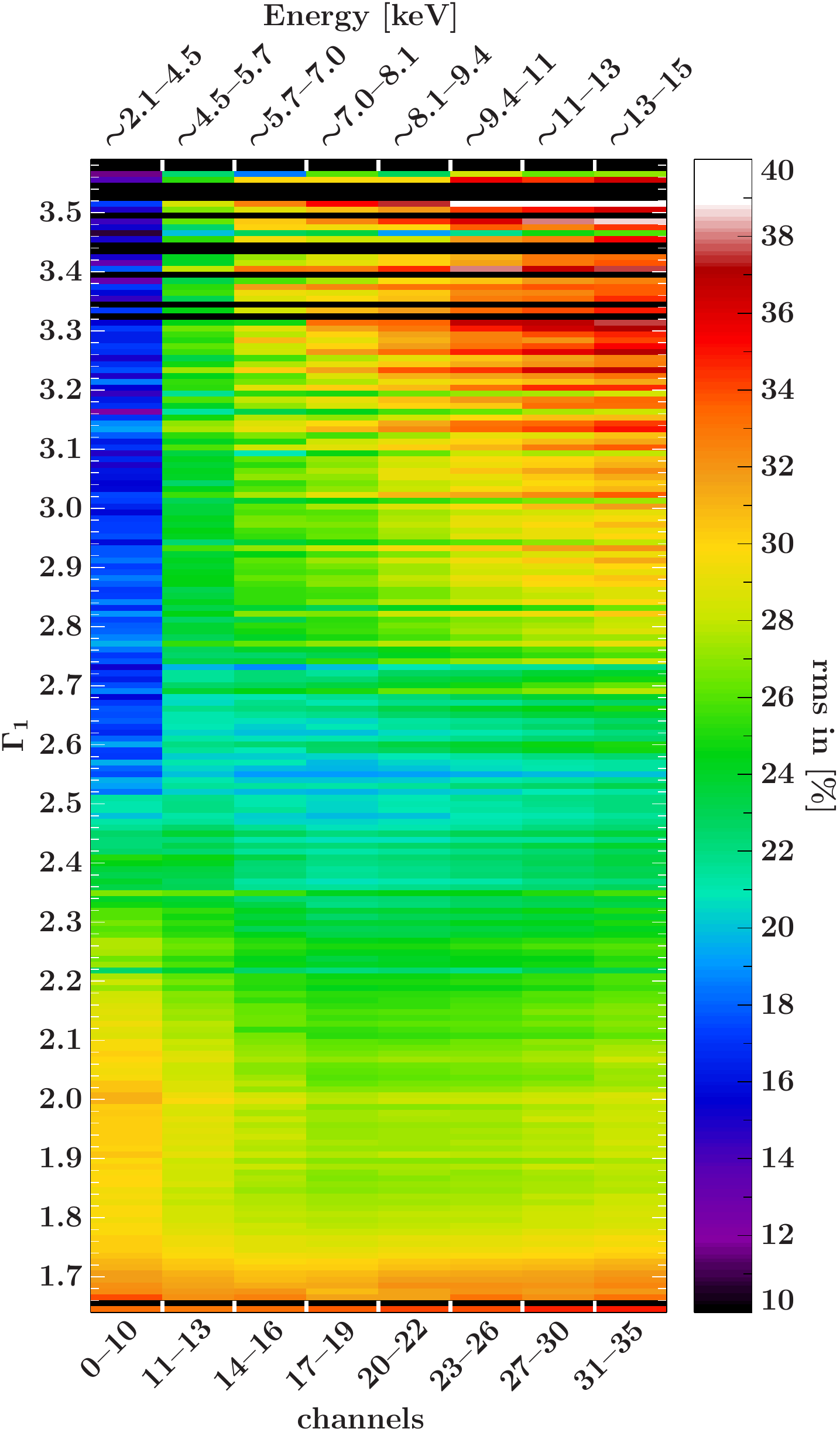}}
  \caption{Evolution of fractional rms between 0.125--256\,Hz in the
    eight binned channels of the \texttt{B\_2ms\_8B\_0\_35\_Q} mode
    with the \mbox{X-ray} spectral shape represented by the soft photon index
    $\Gamma_1$. Black horizontal lines are photon indices for which no
    data are available.}\label{fig:rms_spec}
\end{figure}

To understand the behavior of rms with both energy and spectral shape
better, we look at the behavior of these quantities in the eight bands
of the \verb|B_2ms_8B_0_35_Q| mode (Fig.~\ref{fig:rms_spec}). For each
energy band, we determine the average fractional rms from all
observations during which a similar photon spectral shape was found.
Two photon spectra are considered similar if their respective
$\Gamma_1$ values fall into the same bin of a linear grid with steps
of $\Delta \Gamma_1 = 0.01$ between $\Gamma_{1,\mathrm{min}} = 1.64$
and $\Gamma_{1,\mathrm{max}} = 3.59$. The trends shown in
Fig.~\ref{fig:gamma_rms} and Table~\ref{tab:rmsg1} are confirmed. The
higher energy resolution allows us to see the gradual increase of
fractional rms with $\Gamma_1$ for $\Gamma_1 \gtrsim 2.7$ at all
energies above 4.5\,keV. The increase steepens at higher energies. For
$1.8< \Gamma_1 < 2.3$ the variability is lower above 4.5\,keV than
below that energy, but the decrease in rms with energy does not appear
smooth and shows an indication of recovery above 11\,keV.

Examples of rms spectra shown by \citet{Gierlinski_2010a} agree with
Fig.~\ref{fig:rms_spec}, but these authors do not observe an increase
of the rms above 11\,keV for $1.8< \Gamma_1 <2.3$. Such an increase is
present (but not pointed out) in the one \mbox{Cyg~X-1} rms spectrum shown by
\citet{Munoz-Darias_2010a}. In the 10--200\,keV range,
\citet{Torii_2011a} show that the fractional rms spectra remain
largely flat in hard and intermediate\footnote{\citet{Torii_2011a}
  speak only of hard states with high ASM count rates; a comparison
  with ASM-HID-based state definition of \citet{Grinberg_2013a} shows
  that these states are intermediate.} states and that the
intermediate state shows less variability than the hard state, i.e.,
the behavior presented here for the 2.1--15\,keV range continues at
higher energies.

\subsection{Evolution of PSDs with spectral shape}\label{sect:psds}

Having discussed the overall source variability by studying the rms,
we now turn to the Fourier frequency dependent variability. For our
analysis of the change of PSD shapes with spectral state, we choose an
approach that does not require us to model the PSDs and thus does not
introduce any assumptions on identification of possible components in
individual PSDs and their evolution. For possible pitfalls of such
assumptions see Sect.~\ref{sect:identification}.

We visualize the evolution of the PSD components as a color map in
$f_i$-$\Gamma_1$-space following an idea of \citet{Boeck_2011a}. As in
the analysis of the rms behavior we define a grid with steps of
$\Delta \Gamma_1 = 0.01$ between $\Gamma_{1,\mathrm{min}} = 1.64$ and
$\Gamma_{1,\mathrm{max}} = 3.59$ and 100 equally spaced steps in
logarithmic frequency between 0.125\,Hz and 256\,Hz. When mapping
linear Fourier frequencies to this logarithmic grid, some bins remain
empty. We allocate these to adjacent bins containing at least one
Fourier frequency, equally divided between higher and lower bins. In
the following, we first discuss the energy independent behavior of the
power spectra in Sect.~\ref{sect:total_psd}, and then turn to the
energy dependent variability in
Sect.~\ref{sect:en_psd}.

\subsubsection{Energy-independent evolution of PSDs with
  spectral shape}\label{sect:total_psd}

\begin{figure}
\centering
\resizebox{\hsize}{!}{\includegraphics{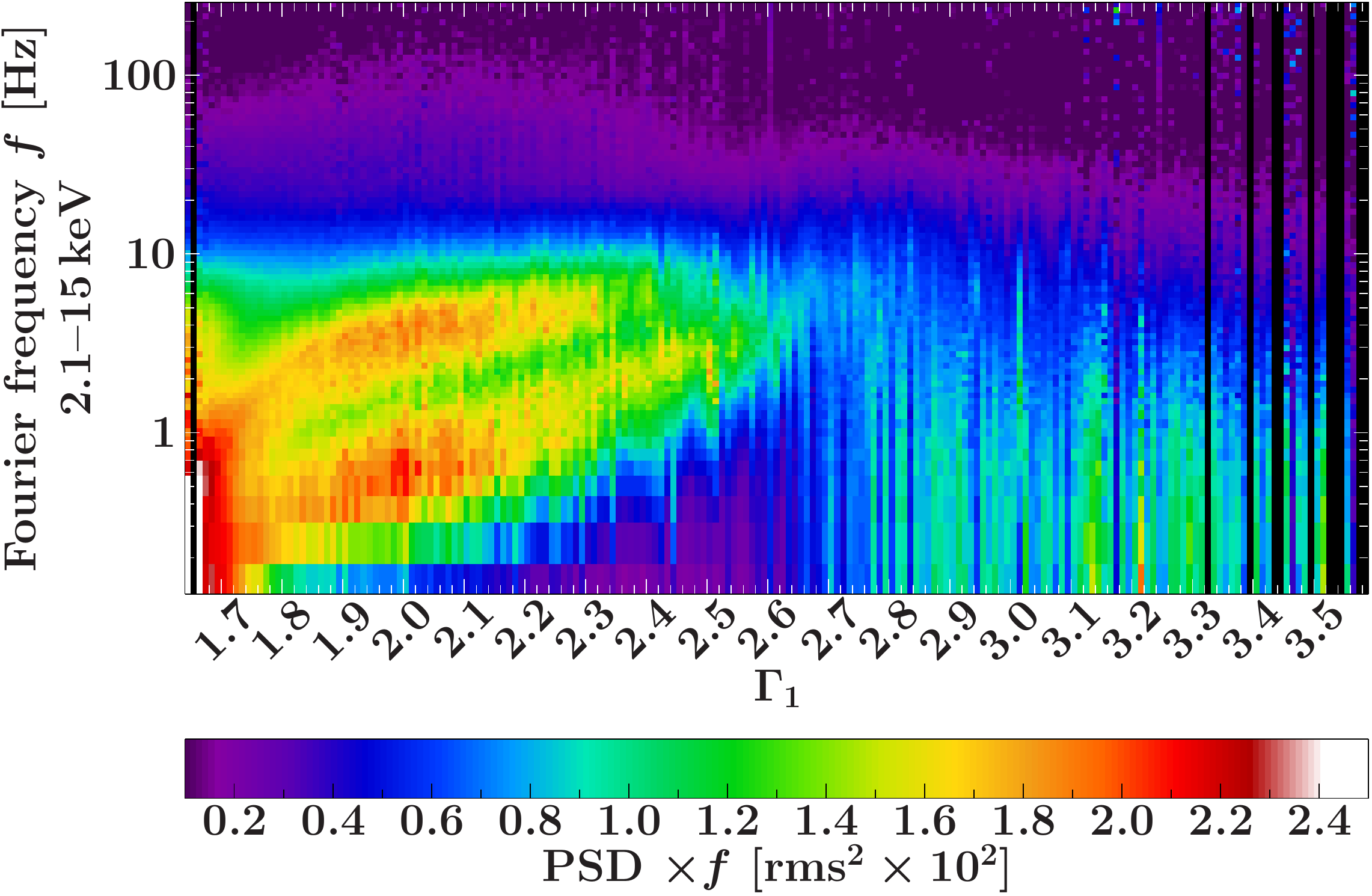}}
\caption{Evolution of PSDs in 2.1--15\,keV band (sum of bands 1--4 and
  full energy range of the \texttt{B\_2ms\_8B\_0\_35\_Q} mode) with
  spectral shape. In this and in all following figures, black vertical
  lines are photon indices for which no data are
  available.}\label{fig:totalpsd}
\end{figure}

Most previous works address the timing quantities in an
energy-independent way. Because of the strong changes in spectral
shape \citep[see][for examples of broadband spectra of \mbox{Cyg~X-1}
in hard and soft states]{Nowak_2012a} and hence count rate ratios at
different energies, extrapolation from energy-band dependent
quantities to the full energy range across all spectral states is
non-trivial. Thus, we first address the full 2.1--15\,keV range and
show the evolution of these PSDs with spectral shape in
Fig.~\ref{fig:totalpsd}.

For $1.75 < \Gamma \lesssim 2.7$, two strong variability components
are present. Previous work has shown that such components can be well
modeled as (often zero-centered) Lorentzians
\citep[e.g.,][]{Nowak_2000a}, and by visual comparison we can identify
the components with the two Lorentzians of \citet{Boeck_2011a} and
with the two lowest frequency Lorentzian components of
\citet{Pottschmidt_2003b}. In the spirit of the non-model dependent
approach chosen here, we will label the lower frequency variability
component as component~1 and the higher frequency variability
component as component~2. Both components shift to higher frequencies
as $\Gamma_1$ softens \citep[see also][]{Cui_1997a,Gilfanov_1999a,
  Pottschmidt_2003b,Axelsson_2005a,Shaposhnikov_2006a,Boeck_2011a}.
The two components appear to be of roughly similar strength. In the
PSDs shown in the 2.1--15\,keV band, we observe that component~2
disappears at $\Gamma_1 \approx 2.5$, so that the PSD is dominated by
component~1 for $2.5< \Gamma_1<2.7$.  Note also the low variability at
the lowest frequencies in the $\Gamma_1 \sim 2.3$--2.6 range.  At the
highest frequencies, the PSDs in all $\Gamma_1$-ranges are dominated
by low statistics and therefore noise.

For $\Gamma_1 \lesssim 1.75$, we see additional power at frequencies
above $\sim$2\,Hz, i.e., above the frequencies assumed by components~1
and~2 in this $\Gamma_1$-range. This behavior is consistent with the
appearance of the third Lorentzian component in the very hard
observations analyzed by \citet{Pottschmidt_2003b} and
\citet{Axelsson_2005a}. We note that the hardest observations analyzed
by these authors are from before 1998 April, i.e., during a time not
covered by the data mode we use, while the hardest spectra analyzed
here were observed during the long hard state between mid-2006 and
mid-2010 \citep{Nowak_2011a,Grinberg_2013a}. Interestingly, the
overall variability amplitude increases in this $\Gamma_1$-range at
all frequencies below $\sim$10\,Hz.

For $\Gamma_1 \gtrsim 2.7$, the variability is generally low,
consistent with low total rms in this range (Fig.~\ref{fig:totalrms}).
It is continuous, without pronounced components, and strongest at the
lowest frequencies and decreases towards higher frequencies.

\subsubsection{Energy-dependent evolution of PSDs with
  spectral shape}\label{sect:en_psd}

\begin{figure}
\resizebox{\hsize}{!}{\includegraphics{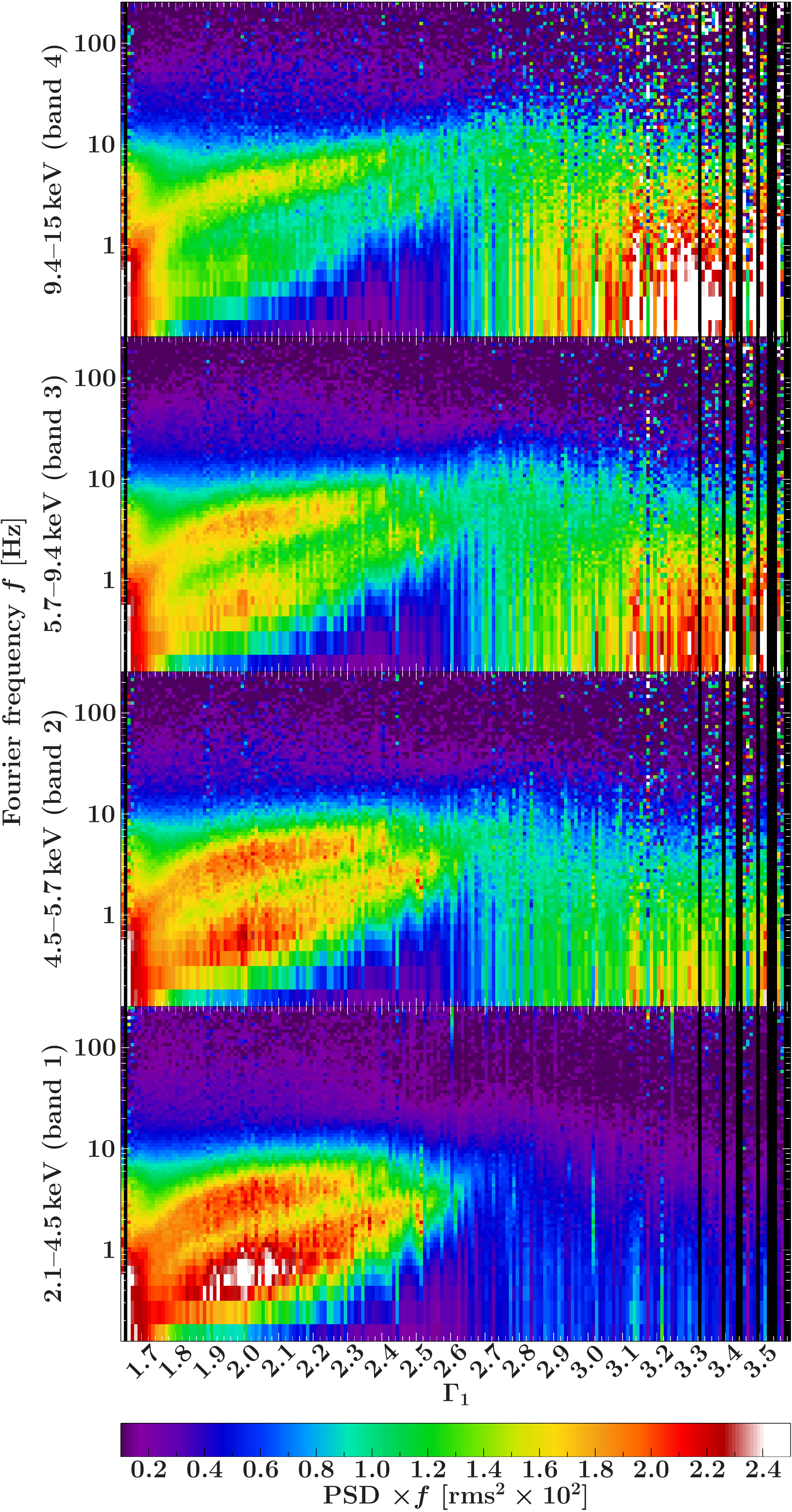}}
\caption{Evolution of the PSDs with spectral shape represented by the
 soft photon index $\Gamma_1$ of the broken power law fit in four
 energy bands (see Table~\ref{tab:bands}). The color scale represents
 averaged $\mathrm{PSD}(f_i) \times f_i$ values at individual Fourier
 frequencies $f_i$.}\label{fig:psd_grid}
\end{figure}

The energy-dependent evolution of the fractional rms
(Sect.~\ref{sect:rms}) hints at the PSDs being highly dependent on the
energy band considered, especially in the soft spectral state. The
shape of the PSDs in all four energy bands is shown in
Fig.~\ref{fig:psd_grid}. These maps emphasize the dominant components
and use a linear color scale for the $\mathrm{PSD}(f_i)\times f_i$
values, while usually individual power spectra are presented on a
logarithmic $\mathrm{PSD}(f_i)\times f_i$-axis
(Figs.~\ref{fig:avg_psds1}, \ref{fig:avg_psds2}, \ref{fig:avg_psds3},
and~\ref{fig:avg_psds4}).

For $1.75 \lesssim \Gamma_1 \lesssim 2.7$, we observe the same two
variability components as in the energy-independent PSDs. Both
components become weaker with increasing energy. The amplitude of
component 1 decreases faster than the amplitude of component 2:
component 1 is stronger than component 2 in band 1 (4.5--5.7\,keV),
but in band 4 (9.4--15\,keV) component 2 is more pronounced.
Especially in the $\Gamma_1 \approx 2.4$--2.6 range, component~1
dominates the lower energy bands, but component~2 the higher energy
bands.
 
For $\Gamma_1 \lesssim 1.75$, the shape of the PSDs appears
energy-independent, as opposed to the softer observations. The
additional higher frequency component discussed in
Sect.~\ref{sect:total_psd} is present in all bands.

For $\Gamma_1 \gtrsim 2.7$, the variability shows a behavior
fundamentally different from the behavior of the two components in the
$1.75 \lesssim \Gamma_1 \lesssim 2.7$ range, as already indicated by
the different behavior of the rms in Sect.~\ref{sect:rms}. In band~1,
the overall variability is low, consistent with the evolution of the
total rms. What variability there is, seems to show a slight trend
towards lower frequencies with increasing $\Gamma_1$ (see also
Fig.~\ref{fig:avg_psds3}). In bands~2 to~4, the variability increases
strongly with increasing energy of the band and within an energy band
with increasing $\Gamma_1$.

\begin{figure}
\resizebox{\hsize}{!}{\includegraphics{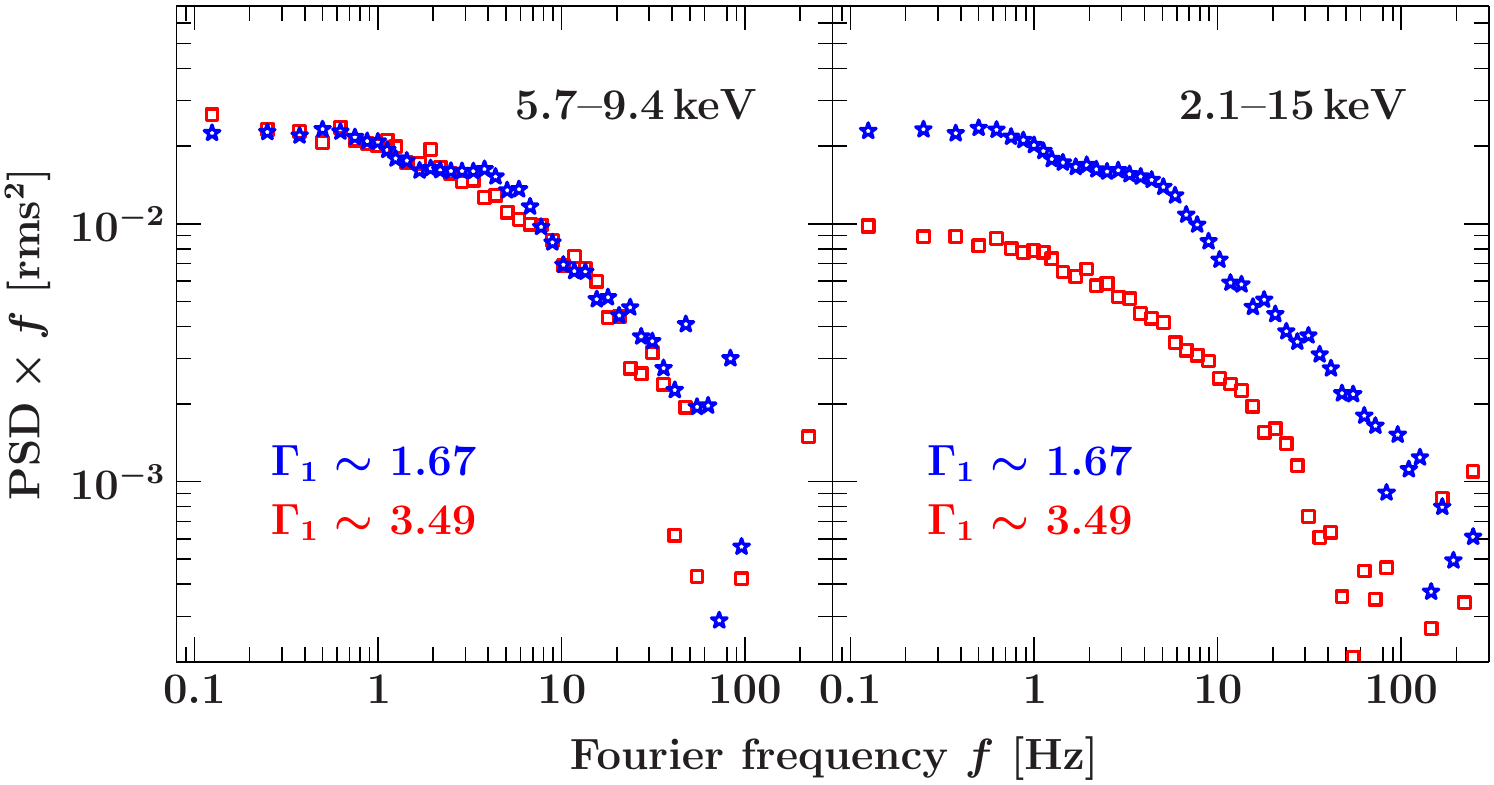}}
\caption{Comparison of average PSDs for $\Gamma_1 \sim 1.67$ (blue
  stars) and $\Gamma_1 \sim 3.49$ (red squares).  PSDs are calculated
  for a logarithmically binned grid with $\mathrm{d} f/f = 0.15$. Each
  PSD is the average of all $n$ PSDs falling within the $\Gamma_1 \pm
  0.01$ interval for the given $\Gamma_1$ values. \textsl{Left}: band
  2 (5.7--9.4\,keV). \textsl{Right}: total band
  (2.1--15\,keV).}\label{fig:avg_psd_comp}
\end{figure}

Figures~\ref{fig:avg_psds1}, \ref{fig:avg_psds2}, \ref{fig:avg_psds3},
and~\ref{fig:avg_psds4} in the appendix present average PSDs at
different spectral shapes to facilitate comparison with previous work.
While the continuous change in the PSD is better illustrated in the
approach of Fig.~\ref{fig:psd_grid}, two effects are better
represented by looking at $f_i$-$\mathrm{PSD}(f_i)\times f_i$-plots of
PSDs. First, in the frequency range considered here, the power spectra
at energies above $\sim$5\,keV are similar for the hardest
observations with $\Gamma_1 < 1.75$ and the softest observations with
$\Gamma_1 > 3.15$, as can be easily seen on
Figs.~\ref{fig:avg_psds1}--\ref{fig:avg_psds4} and as shown for two
example $\Gamma_1$ values in Fig.~\ref{fig:avg_psd_comp}. While the
different dependence on the energy and the different cross-spectral
properties (see Sect.~\ref{sect:cross}) suggest a different origin for
these power spectra, the shapes can easily be confused, especially
when data are of lower quality. Secondly, there are indications of a
weak third hump above 10\,Hz for $1.75 \lesssim \Gamma_1 \lesssim
2.15$ that is not visible in this range in Fig.~\ref{fig:psd_grid} and
that corresponds to the higher frequency Lorentzian component $L_3$
modeled by \citet{Pottschmidt_2003b}. For the hardest observations
with $\Gamma_1 \lesssim 1.75$ there are also hints of a further
component at $\sim$30\,keV \citep[see
also][]{Revnivtsev_2000a,Nowak_2000a,Pottschmidt_2003b}.

An energy-dependent approach to both rms and PSDs is clearly necessary
to understand the variability components in different states as they
show strikingly different behavior with changing energy that is
otherwise missed. This result may be especially important when
comparing black hole binaries to AGN and cataclysmic variables, where
at the same energy we probe different parts of the emission (see also
Sect.~\ref{sect:agn}).

\subsubsection{Discussion of PSD shapes}\label{sect:psd_comp}

The frequency shift of the variability components of the PSDs in the
RXTE range with the softening of the source in the hard and
intermediate state continues at higher energies
\citep{Cui_1997a,Pottschmidt_2006a,Cabanac_2011a} up to $\sim$200\,keV
\citep{Torii_2011a}, even though such high energy analyses are rare.
The different energy dependence of the components has been previously
noted for individual observations, both for the presented energy range
and above \citep{Cui_1997a,Nowak_1999a,Pottschmidt_2006a,Boeck_2011a},
but never shown for such a comprehensive data set as presented in this
work.

The $\sim$$f^{-1}$-variability of the soft state PSDs can be described
by a power law, while the power spectra of intermediate states are
often modeled with a combination of Lorentzian components with
(cut-off) power law \citep[e.g.,][]{Cui_1997a,
  Axelsson_2005a,Axelsson_2006a}. \citet{Cui_1997a} have noted that
the power law component becomes stronger with increasing energy for
individual soft state observations up to 60\,keV; we observe this
effect, which has also been noted by \citet{Churazov_2001a}, up to our
maximum energy of 15\,keV. We also clearly see that at energies above
4.5\,keV the variability grows with increasing $\Gamma_1$ for
$\Gamma_1 > 2.7$, but decreases below 4.5\,keV.

Our approach does not allow us to track the weak power law in the
intermediate state, especially given that our lowest accessible
frequency is 0.125\,Hz. However, we see that the power law dominates
the PSDs for $\Gamma_1 \gtrsim 2.7$ -- the abrupt change can also be
seen in the cross spectral quantities presented in the next Section
(Sect.~\ref{sect:cross}). In particular, for $\Gamma_1 \gtrsim 2.7$,
the coherence, which shows a dip for $2.5 < \Gamma_1 < 2.7$, recovers
(Sect.~\ref{sect:cof}); the time lag spectra lack the previously
present structure (Sect.~\ref{sect:lagspec}); and the averaged time
lags show an abrupt drop and no correlation with $\Gamma_1$ for
$\Gamma_1 \gtrsim 2.65$ (Sect.~\ref{sect:avglags}). Cross-spectral
quantities are thus clearly crucial to better understand the
variability of the source and are discussed in the following Section.

\section{Coherence and time lags}\label{sect:cross}

\subsection{Evolution of Fourier-dependent coherence with spectral
  shape}\label{sect:cof}

\begin{figure}
\centering
\resizebox{\hsize}{!}{\includegraphics{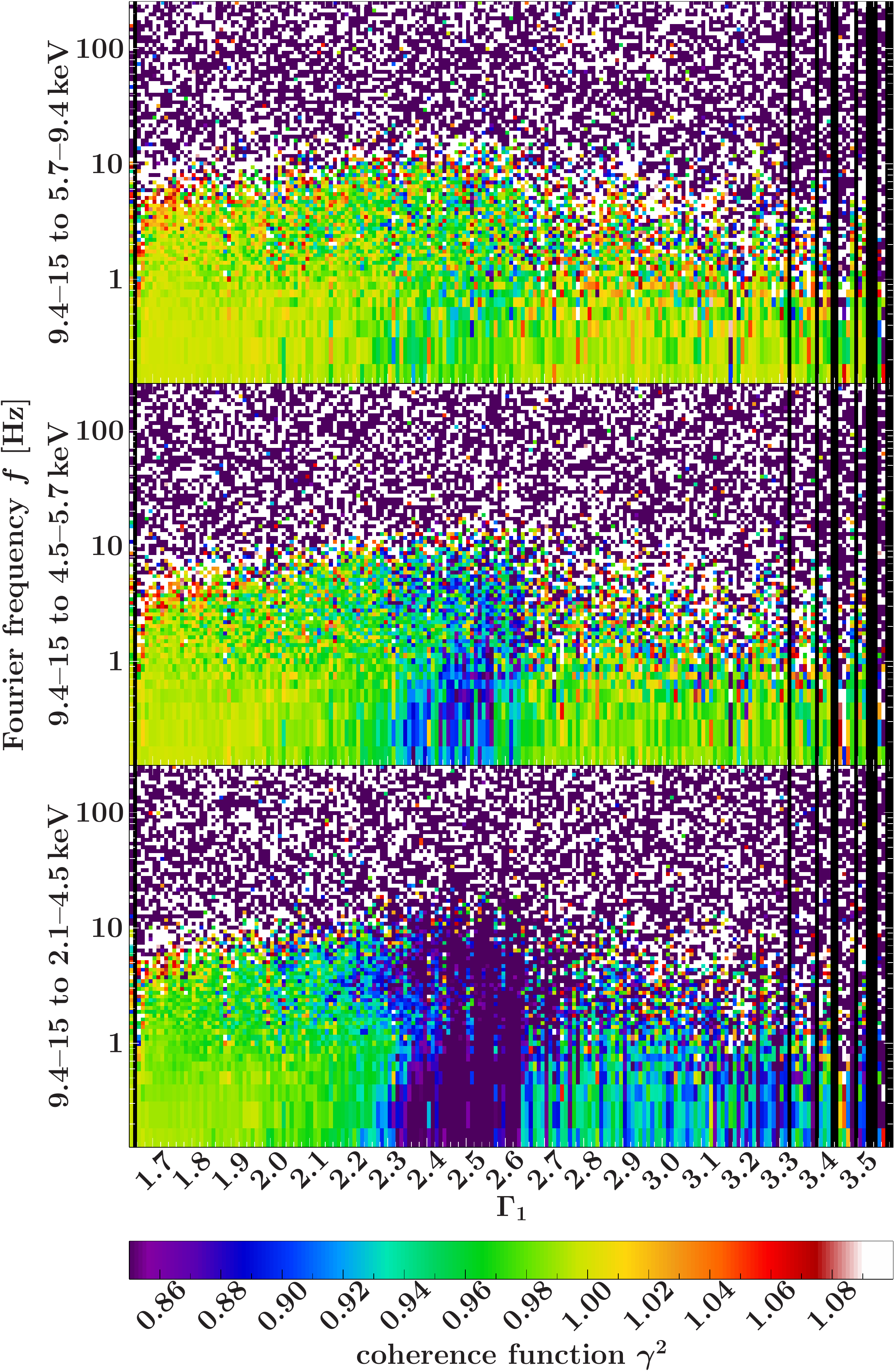}}
\caption{Evolution of the coherence, $\gamma^2$, with spectral shape
  represented by the soft photon index $\Gamma_1$ of the broken power
  law. The color scale represents averaged $\gamma^2$ values at
  individual Fourier frequencies $f_i$. \textsl{Upper panel:}
  $\gamma^2_{4,3}$ between band 4 and 3; \textsl{middle panel:}
  $\gamma^2_{4,2}$ between bands 4 and 2, \textsl{lower panel:}
  $\gamma^2_{4,1}$ between bands 4 and 1.}\label{fig:cof_grid}
\end{figure}

We calculate maps of the coherence function, $\gamma^2_{h,s} (f_i)$
where $h \in \{4,3,2\}$ is the harder band and $s \in \{3,2,1\}$ is
the respective softer band, using the same grid as in our analysis of
the power spectra (Sect.~\ref{sect:psds}). Figure~\ref{fig:cof_grid}
shows maps of the coherence function between band 4 and bands 1, 2,
and 3. By eye, we can identify an envelope in all shown coherence
functions that follows the high frequency outline of the dominant
variability components of the power spectra
(Sect.~\ref{sect:psds}). Outside of this envelope, the coherence
fluctuates strongly.

To assess whether the envelope is an intrinsic feature of the source
or whether it is due to the low signal at the high frequencies, we
simulate example light curves with typical spectra and PSDs for
different values of $\Gamma_1$. To do so, we use Monte Carlo
simulation with the software package SIXTE (SImulation of \mbox{X-ray}
TElescopes) developed for the analysis of various \mbox{X-ray} instruments
\citep{Schmid_2012_PhD}. For each selected PSD provided in the SIMPUT
(SIMulation inPUT) file
format\footnote{\url{http://hea-www.harvard.edu/HEASARC/formats/simput-1.0.0.pdf}},
the simulation software obtains a light curve using the algorithm of
\citet{Timmer_1995a}. Here, we use the PSDs in the full 2.1--15\,keV
band as input PSDs. Based on this light curve, a sample of
\mbox{X-ray} photons is generated with the Poisson arrival process
generator of \citet{Klein_1984a}. The energies of the photons are
distributed according to the observed spectrum, which is chosen as
constant throughout the simulation. The photons are then processed
through a model of the PCA taking into account instrumental effects
such as energy resolution and dead time. The output of this model is a
list of simulated events as detected with the PCA, which is converted
back to light curves in chosen channel or energy bands for the
subsequent analysis. As the spectral shape remains unchanged and only
the rate of generated photons, i.e., the normalization of the
spectrum, varies with the light curve, the variability in different
energy bands is, per definition, coherent.

We simulate light curves in band 4 and band 1 and calculate
$\gamma^2_{4,1} (f_i)$ as for observed light curves. The trends we
observe in $\gamma^2_{4,1} (f_i)$ are the same as for observed light
curves: the coherence function is $\sim$1 below approximately 10\,Hz
and then shows strong variations above that frequency. This threshold
frequency is weakly dependent on state and decreases towards lower
frequencies in the soft state. Hence, the envelope seen in the
coherence function plots is not source-intrinsic but a consequence of
the low signal at higher frequencies.

A feature not present in our simulations of ideally coherent light
curves is the decrease in coherence for $2.35 \lesssim \Gamma_1
\lesssim 2.65$ in a region which seems to follow the shape of the PSD
components~1 and~2 (see Sect.~\ref{sect:total_psd}). This decrease is
most evident in $\gamma^2_{4,1}(f_i)$, where the coherence values drop
below 0.85, but is also well visible in $\gamma^2_{4,2}(f_i)$ and
indicated in $\gamma^2_{4,3}(f_i)$ (Fig.~\ref{fig:cof_grid}).

For $\Gamma_1 \gtrsim 2.65$, the coherence recovers, although the
values between the 9.4--15\,keV and 2.5--4.5\,keV bands remain
significantly lower ($\sim$0.92--0.94) than between 9.4--15\,keV and
the 4.5--5.7\,keV and 5.7--9.4\,keV bands. The coherence seems also
noisier, although this is likely due to the lower number of
observations in the soft state.

\subsection{Evolution of Fourier-dependent time lags with spectral
  shape}\label{sect:lagspec}

\begin{figure}
\resizebox{\hsize}{!}{\includegraphics{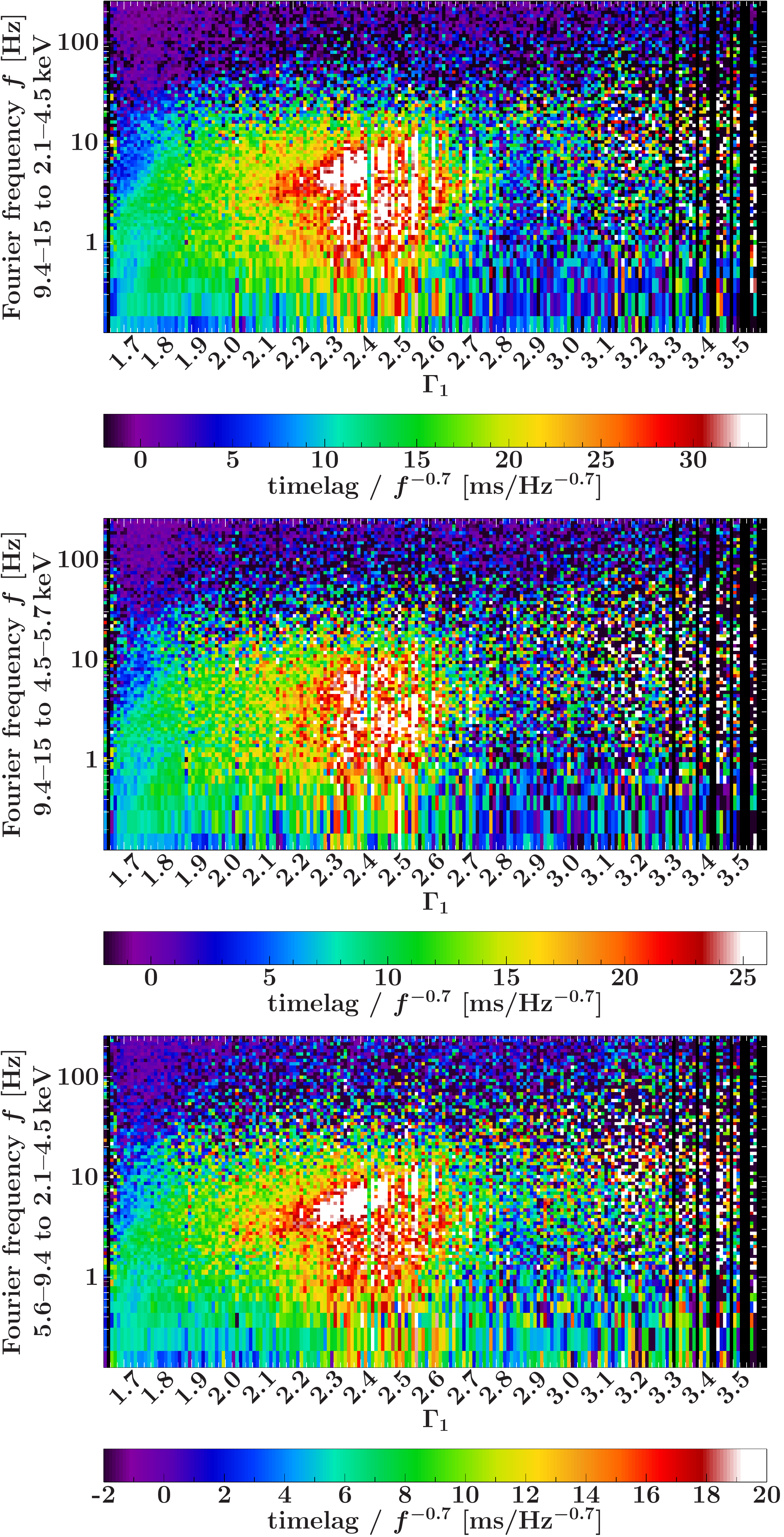}}
\caption{Evolution of time lags $\delta t$ in the $ \delta t (f_i) /
  f_i^{-0.7} \coloneqq \Delta \delta t$ representation with spectral
  shape represented by the soft photon index $\Gamma_1$ of the broken
  power law. The color scale (note the different scales for the
  presented bands) represents averaged $\Delta \delta t$ values at
  individual Fourier frequencies $f_i$. Positive lags means that the
  hard photons lags the soft.}\label{fig:lag_grid}
\end{figure}

The time lags show a strong power law dependence on the Fourier
frequency with $\delta t (f_i) \propto f_i^{-0.7}$ \citep[][and
references therein]{Nowak_1999a}, so that we consider the fraction
$\delta t (f_i) / f_i^{-0.7} \coloneqq \Delta \delta t$ to visualize
structures in the time lag spectra and again follow the approach from
Sect.~\ref{sect:psds}.  In Fig.~\ref{fig:lag_grid}, we present maps of
$\Delta \delta t$ for band combinations showing the largest lags,
i.e., bands with largest separation in energy
\citep{Miyamoto_1988a,Nowak_1999a}. In the phase lag representation,
the largest lags between the 2.1--4.5\,keV and 9.4--15\,keV bands do
not exceed $\sim$0.5\,rad, so that the time lags are well defined and
the features in Fig.~\ref{fig:lag_grid} are not due to the phase lag
to time lag conversion (see Sect.~\ref{sect:timing}).

If the lags strictly followed $\delta t (f_i) \propto f_i^{-0.7}$,
then the only structure $\Delta\delta t$ would show would be a
gradient with changing $\Gamma_1$, corresponding to the known changes
in average lag with state \citep[e.g.,][]{Pottschmidt_2003b}. Such a
gradient is indeed visible, with the lags obtaining highest values at
$\Gamma_1 \sim 2.5$--2.6.

For $\Gamma_1 \lesssim 2.7$, however, there is a $f_i$-dependent
structure overlaid onto the $\Gamma_1$-dependent gradient.  This
structure seems to track the two components~1 and~2 of the PSDs (see
Sect.~\ref{sect:psds}) and is visible for all energy band
combinations.  A correlation between the features of the PSDs and of
time lag spectra has been suggested in \mbox{Cyg~X-1} by
\citet{Nowak_2000a}, but to our knowledge has not been shown for a
wide range of spectral states and therefore PSD shapes for any black
hole binary previously.

For $\Gamma_1 \gtrsim 2.7 $, the lag spectrum shows no structure and
is has smaller values than for harder $\Gamma_1$. No evolution with
$\Gamma_1$ is seen in the soft state.

For $\Gamma_1 < 1.75$, where the PSDs show increased variability and a
possible additional component at higher frequencies, no corresponding
changes can be seen in the $\Delta \delta t$ maps. The two
components~1 and~2 of the hard and intermediate state PSDs are,
however, clearly visible in the time lag maps.

\subsection{Evolution of average time lag with spectral
  shape}\label{sect:avglags}

\begin{figure}
\resizebox{\hsize}{!}{\includegraphics{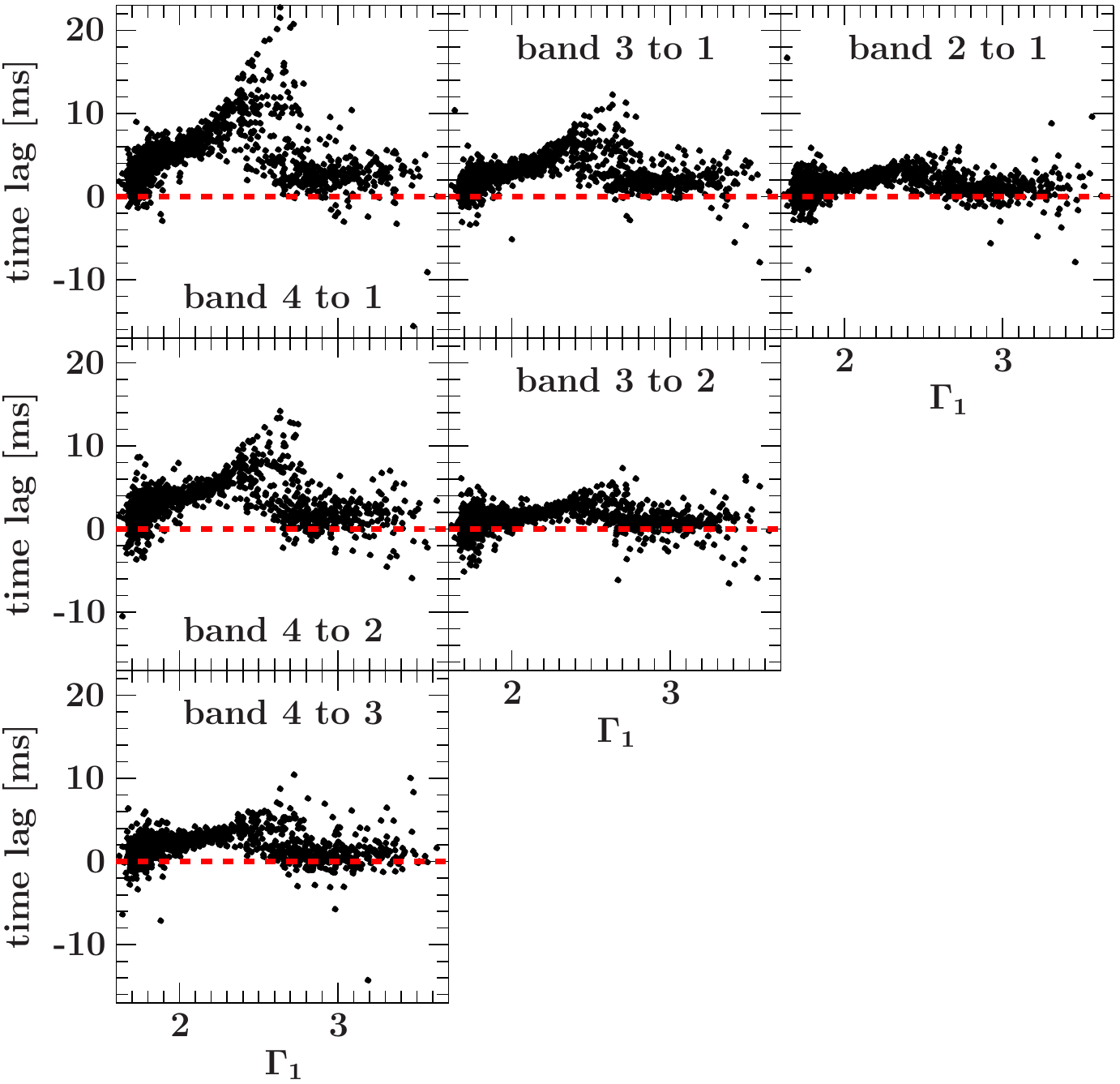}}
\caption{Averaged time lag in the 3.2--10\,Hz range vs.\ soft photon
  index of the broken power law fits $\Gamma_1$ for all combinations
  of energy bands (band 1: 2.1--4.5\,keV, band 2: 4.5--5.7\,keV, band
  3: 5.7--9.4\,keV, band 4: 9.4--15\,keV). Zero time lag is
  represented by the red dashed line.}\label{fig:gamma_lag}
\end{figure}

\begin{table*}
  \caption{Spearman correlations coefficients $\rho$ and null hypothesis 
    probabilities $P$ for the correlation of averaged time lags in the
    3.2--10\,Hz band with  $\Gamma_1$ for different $\Gamma_1$ ranges
    (see also Fig.~\ref{fig:gamma_lag}).}\label{tab:lagg1} 
  \centering
\begin{tabular}{lcccccccccccc}
\hline
\hline
&\multicolumn{2}{c}{band 4 to 1} & \multicolumn{2}{c}{band 4 to 2}& \multicolumn{2}{c}{band 4 to 3}& \multicolumn{2}{c}{band 3 to  1}& \multicolumn{2}{c}{band 3 to 2}& \multicolumn{2}{c}{band 2 to 1}\\
&$\rho$&$P$&$\rho$&$P$&$\rho$&$P$&$\rho$&$P$&$\rho$&$P$&$\rho$&$P$\\
\hline
$\Gamma_1 < 2.65$&0.79&0\tablefootmark{a}&0.72&0\tablefootmark{a}&0.65&0\tablefootmark{a}&0.3&0\tablefootmark{a}&0.52&0\tablefootmark{a}&0.52&0\tablefootmark{a}\\
$\Gamma_1 \geq 2.65$&$-0.14$&0.01&$-0.12$&0.03&$-0.06$&0.30&$-0.09$&0.10&$-0.10$&0.06&$-0.09$&0.11\\
\hline
\end{tabular}
\tablefoot{\tablefoottext{a}{within numerical accuracy}}
\end{table*}

The usual measure for time lags in the literature are averaged lags
$\delta t_\mathrm{avg}$ in the 3.2--10\,Hz band
\citep[e.g.,][]{Pottschmidt_2003b,Boeck_2011a}. To facilitate
comparisons, we follow the approach of the previous work: we
rebin each raw time lag spectrum to a logarithmically spaced frequency
grid with $\mathrm{d}f/f = 0.15$ and then calculate an average value
for the 3.2--10\,Hz band.

The average time lags (Fig.~\ref{fig:gamma_lag}) are larger for bands
with a greater difference in energy, but all combinations of energy
bands show the same non-linear trends with $\Gamma_1$: a strong
increase of the time lags from hard to intermediate states and an
abrupt drop at $\Gamma_1 \sim 2.65$ that has also been noted by, e.g.,
\citet{Pottschmidt_2000a} and \citet{Boeck_2011a}.

Spearman rank coefficients between the average time lag, $\delta
t_\mathrm{avg}$, and the soft photon index, $\Gamma_1$, for $\Gamma_1
< 2.65$ and $\Gamma_1 \geq 2.65$ are listed in
Table~\ref{tab:lagg1}. In all bands, $\delta t_\mathrm{avg}$ is
strongly positively correlated with $\Gamma_1$ for $\Gamma_1 < 2.65$
and not correlated or weakly negatively correlated with $\Gamma_1$ for
$\Gamma_1 \geq 2.65$. The weak negative correlation may be an artifact
of the choice of threshold $\Gamma_1$-values of 2.65.  Spearman rank
coefficients and their low null hypothesis probabilities do, however,
not imply a linear correlation: the relationship for $\Gamma_1 < 2.65$
appears non-linear, with a bend at $\Gamma_1 \sim1.8$--1.9 and a
stronger than linear increase towards higher $\Gamma_1$
(Fig.~\ref{fig:gamma_lag}, especially time lag between band~4 and
band~1).

\subsection{Comparison with previous results for coherence and lags}

Cross-spectral analysis is generally less present in the literature
and an analysis of the Fourier-dependent coherence function and time
lags for \mbox{Cyg~X-1} across the full range of spectral states of
the source is undertaken in this paper for the first time. However,
the available previous results agree well with the long-term evolution
presented here, in particular for coherence in different spectral
states \citep{Cui_1997a,Pottschmidt_2003b,Boeck_2011a} and for average
time lags in hard and intermediate states \citep{Pottschmidt_2003b,
  Boeck_2011a}.  The return to low values in the soft state has been
noted before \citep{Pottschmidt_2000a,Boeck_2011a}, but the abruptness
of the change with $\Gamma_1$ can only be seen for an extensive data
set such as the one presented here or for the extremely rare, almost
uninterruptedly covered transitions such as the one analyzed by
\citet{Boeck_2011a}.

In a different approach, \citet{Skipper_2013a} compute the cross
correlation function between the photon index and the 3--20\,keV count
rate of \mbox{Cyg~X-1} and find different shapes in dim hard states,
bright hard states, and soft states. All their hard state cross
correlation functions show asymmetries that can only be explained if a
part of the hard count rate lags the soft, consistent with our
observations of a hard lag. Their dim hard states correspond to the
hardest states observed by \citet{Pottschmidt_2003b} in 1998. These
hardest states are similar to our data falling into the $\Gamma_1 <
1.8$ range, i.e., mostly data from the long hard state 2006 to 2010
\citep{Nowak_2011a,Grinberg_2013a}, and therefore before the first
bend of the $\delta t_\mathrm{avg}$-$\Gamma_1$ correlation
(Sect.~\ref{sect:avglags}). Data taken during these states include
further components in the PSD, which can also be seen in the hardest
observations presented here. Similar evidence for growing hard lags as
the source transits into the intermediate state was also found by
\citet{Torii_2011a} in cross correlation function of the 10--60\,keV
and 60--200\,keV \textsl{Suzaku} light curves.

The double-humped structure in the phase lag spectra of \mbox{Cyg~X-1}
(Sect.~\ref{sect:lagspec}, for $\Gamma_1 \lesssim 2.7$) has been
suggested since the late 1980s \citep{Miyamoto_1989a}. Correlated
features in the PSDs and time lag spectra were noted for individual
observations \citep{Cui_1997a,Nowak_2000a}, but there has only been
little work on the systematic evolution of the features with spectral
state. We stress the importance of this approach by comparing our
results to \citet{Pottschmidt_2000a}, who discuss time lag spectra for
individual observations in different state and argue on their basis
for a similar shape of the time lag spectra in hard and soft states.
While the $f^{-0.7}$-behavior is indeed dominant in both states and
individual hard and soft observations therefore appear similar, we see
features correlated with PSD components in the hard but not in the
soft state (Fig.~\ref{fig:lag_grid}).

\section{Using the variability of \mbox{Cyg~X-1} as a template for
  other black holes}\label{sect:other}

Before discussing possible implications of the observed variability
patterns on physical interpretations of the variability in
Sect.~\ref{sect:theory}, we first address the importance of our
model-independent approach for long-term timing analysis and the
implications for the interpretation and further analysis of
variability components in other sources.

\subsection{\mbox{Cyg~X-1} and the canonical picture of states and
  state transitions}\label{sect:canon_bh}

A comprehensive analysis of all sources observed with RXTE is out of
the scope of this work. In contrast to most transient black hole
binaries \citep[see, e.g., ][for a sample study]{Klein-Wolt_2008a} and
most notably the canonical \mbox{X-ray} transient \object{GX 339$-$4}
\citep{Belloni_2005a}, \mbox{Cyg~X-1} does not show strong narrow quasi
periodic oscillations (QPOs)\footnote{Here and in the remainder of
  this work we use the term ``QPO'' exclusively for narrow features
  and do not use the term to describe the broader humps as done, e.g.,
  by \citealt{Shaposhnikov_2006a}. The difference in terminology is
  phenomenological and does not necessarily imply an assumption of
  different origin of the features.}, although there is evidence for
short lived QPOs \citep[][R.~Remillard, priv.\
comm.]{Pottschmidt_2003b}. Common for black hole binaries, however,
seem to be the two flavors of the hard state with different \mbox{X-ray}
timing properties \citep[e.g.,][for Swift J1753.5$-$0127 and
\citealt{Done_2005a}, for \object{XTE~J1550$-$564}]{Cassatella_2012b},
the frequency shift of the main variability components in the hard and
intermediate states \citep[e.g,][]{Klein-Wolt_2008a}, and the sharp
change in timing behavior as the source transits into the soft state
\citep[e.g.,][]{Nowak_1995,Klein-Wolt_2008a}, all shown here in
unprecedented detail for \mbox{Cyg~X-1}.

In the canonical picture of the black hole binary states, the rms
drops to a few percent in the soft state \citep{Belloni_2010a}. The
rms of \mbox{Cyg~X-1} does not assume such low values, even when the
2.1--4.5\,keV band is considered (Fig.~\ref{fig:totalrms}). However,
\mbox{Cyg~X-1} is not the only source showing such behavior: for
example, a complex, energy dependent rms behavior in the soft state
with the rms remaining flat in the 3.2--6.1\,keV band and increasing
in the 6.1--10.2\,keV band has been also observed in the 2010 outburst
of \object{MAXI~J1659$-$152} \citep{Munoz-Darias_2011a}, a source
that, like \mbox{Cyg~X-1}, does not show a purely disk-dominated soft
state.  Similar to \mbox{Cyg~X-1}, MAXI~J1659$-$152 also shows
increased time lags in the intermediate state
\citep{Munoz-Darias_2011a}. Correlated features in power and time lag
spectra were also seen, e.g., in GX 339$-$4 \citep{Nowak_2000a} and
Swift J1753.5$-$0127 \citep{Cassatella_2012a}.

\subsection{The identification of the variability
  components}\label{sect:identification}

Assuming that we can use \mbox{Cyg~X-1} as a template for the
long-term variability of PSDs and cross-power quantities in other
black hole binaries, we can avoid mis-identification of PSD components
in the different spectral states. This is often a problem in
observations which are less well sampled than desirable and therefore
do not allow to track the frequency dependency of components in the
PSDs with spectral shape. We note, however, that as a persistent
source \mbox{Cyg~X-1} may not cover some of the extreme behavior of
the transient sources.

For example, in their study of a state transition of
\object{XTE~J1650$-$500}, \citet{Kalemci_2003a} model the power
spectrum in the intermediate state as a Lorentzian, which shifts to a
higher frequency when comparing the power spectra taken from the
3--6\,keV and the 6--15\,keV data. A comparison with
Fig.~\ref{fig:psd_grid} shows that the observations of
\citet{Kalemci_2003a} could be explained if this source falls in the
spectral range where the lower energy bands are dominated by
component~1 and the higher energy bands by component~2 ($2.4 \lesssim
\Gamma_1 \lesssim 2.7$ for \mbox{Cyg~X-1}).

A comparison with Figs.~\ref{fig:totalpsd} and~\ref{fig:psd_grid} also
reveals a different interpretation of the RXTE observations of
Swift~J1753.5$-$0127. Here, \citet{Soleri_2013a} see two broad peaks
which decrease in frequency with increasing spectral hardness, and
then jump again to higher frequencies for the hardest
observations. The overall pattern seen in our \mbox{Cyg~X-1} data
suggests the alternative interpretation that for softer observations
the model of \citet{Soleri_2013a} tracks components~1 and~2, while for
the harder observations the model picks up the prominent component~3
that appears at higher frequencies than components~1 and~2 in our
hardest observations \citep[see also][]{Pottschmidt_2003b}.

\subsection{Variability in black hole binaries and AGN}\label{sect:agn}

Since the physics of accretion is expected to be similar for compact
objects of different mass, the characteristic frequencies seen in the
power spectra are expected to scale with mass. AGN are therefore
expected to show timing behavior similar to black hole binaries,
albeit at higher timescales of hours to years \citep{McHardy_2006a}.
Timing analysis of AGN is, however, notoriously complicated because of
the low fluxes and the low frequencies, which are hard to sample well.

PSDs of AGN generally suffer from lower signal-to-noise and sparser
sampling than those of black hole binaries, and they are usually
modeled with either power laws or broken/bending power laws
\citep[see, e.g.,][ for analysis of a large sample observed with
XMM-Newton]{Gonzalez-Martin_2012a}. More complex model shapes with two
bends (going to a power-law slope of 0 at the lowest temporal
frequencies) or two Lorentzians have been fit only in one case so far,
namely \object{Akn~564}
\citep{McHardy_2007a}\footnote{\citet{Markowitz_2003a} tentatively
  modeled a second break in the PSD of \object{NGC~3783}, but
  \citet{Summons_2007a} re-measured the PSD, finding only one break.}.
AGN generally show high coherence that drops at higher frequencies and
Fourier frequency- and energy-dependent time lags
\citep{Vaughan_2003a,Markowitz_2007a,Sriram_2009a}. In Akn~564, the
time lag spectra show features at frequencies corresponding to changes
in the components dominating the PSD \citep{McHardy_2007a}. On short
time scales, AGN show soft lags between energy bands dominated by soft
excess and the power law \citep[e.g.,][]{Fabian_2009a,De_Marco_2013a},
but note that the mass-scaled frequency range covered by these
observations is not accessible for lag studies in \mbox{X-ray}
binaries\footnote{Typical frequencies probed in AGN using XMM long
  looks that allow for lag studies are in the $10^{-5}$--$10^{-3}$\,Hz
  range, corresponding to roughly 10--1000\,Hz in black hole
  binaries. On the other hand, broadband AGN PSDs are calculated from
  combined long-term lightcurves with XMM and RXTE that probe the
  $10^{-8}$--$10^{-3}$\,Hz range, corresponding to roughly
  $10^{-2}$--1000\,Hz in black hole binaries.}. Although they differ
in details, the correlations between spectral and timing parameters in
AGN and binaries also seem to follow similar trends
\citep{Papadakis_2009a} and appear to be at least comparable.

As we have seen above (Sect.~\ref{sect:en_psd}
and~\ref{sect:canon_bh}), PSD shapes in \mbox{Cyg~X-1} and other black
hole binaries are, however, strongly
energy-dependent. Energy-dependence of AGN PSDs has not been as
well-studied, but relatively higher-energy PSDs tend to show either
higher break frequencies \citep[e.g.,][]{McHardy_2004a,
  Markowitz_2007a} or flatter power-law slopes above the break
\citep{Nandra_2001a,Vaughan_2003a, Markowitz_2005a} with PSD shape
flattening and normalization dropping above 10 keV in the case of
\object{NGC~7469} \citep{Markowitz_2010a}. A simple comparison of
power spectra of black hole binaries and AGN at frequencies corrected
for the different mass but at the same energies may therefore not be
appropriate, in particular since the accretion disk temperatures are
much lower in AGN \citep[see
also][]{McHardy_2004a,Done_2005a,Markowitz_2007a,Papadakis_2009a}.
This is especially important since at the high signal to noise
available here we find that for all presented PSD the overall shape of
the PSDs changes with energy, except for the hardest and softest
observations (Figs.~\ref{fig:avg_psds1}--\ref{fig:avg_psds4}). Using
bands without a significant direct contribution from the disk or
relying on the presence of the disk simply changing the normalization
of the PSD in the observed range between different energy bands
\citep[as is the case deep in the soft state,][]{Churazov_2001a} may
therefore be an oversimplified approach given the complex behavior the
PSDs exhibit.

A further complication implied by our data is the qualitative
similarity between the shape of the hardest and softest spectra in the
considered frequency range
(Figs.~\ref{fig:avg_psds1}--\ref{fig:avg_psds4},
Fig.~\ref{fig:avg_psd_comp}). This may pose a problem when assigning
states to PSDs of AGN, which are more affected by noise and have
usually a poorer temporal frequency coverage than \mbox{Cyg~X-1} and
where typically only one or a few PSDs per source are available, so
that long-term evolution of variability components over several
spectral states cannot be tracked.

\section{Models for \mbox{X-ray} variability}\label{sect:theory}

As shown, e.g., by \citet{Nowak_1999a} and \citet{Nowak_2000a}, the
general shape of the PSDs can be well explained as the sum of multiple
broad Lorentzians. A possible interpretation of these Lorentzians in
the earlier models was that they represented some heavily damped
oscillating structures, which could be located, e.g., in the accretion
disk. \citet{Nowak_2000a} showed that if these Lorentzians are due to
statistically independent physical processes, it is in principle
possible to also reproduce the general shape of the frequency
dependent \mbox{X-ray} lags and of the coherence function.
\citet{Koerding_2004a} extended this approach by showing that a power
law photon spectrum with the photon index varying on short timescales
(a ``pivoting power law'') could reproduce the main black hole
variability characteristics.

Given the fact that with such fairly simple assumptions it is already
possible to generate a large spectrum of variability properties, it is
not surprising that a large number of different physical models for
the variability exist in the literature. These models are often based
on fundamentally different assumptions on how the \mbox{X-ray}
variability is generated. The models also differ in their assumption
of the \mbox{X-ray} emitting geometry, such as the shape of a
Comptonizing corona, the contribution of the jet to the observed
\mbox{X-ray}s, and the truncation of the disk. Physical models that
consistently explain timing and spectral variability are lacking --
even the very source for the general shape of the PSDs is not fully
understood yet. Offering such a model or a detailed review of all
current models is therefore out of the scope of this paper. Our aim is
instead to point out the different approaches in the light of the
observational results presetented in this work, although we exclude
models in which Compton reflection is playing a major part in shaping
the short-term variability, as these have been ruled out in
spectro-timing studies of the black hole binary GX 339$-$4
\citep{Cassatella_2012a}.

A wide class of models for the variability is based on the idea of
inwards propagating mass accretion fluctuations that originate at
certain radii in the disk and vary on the local viscous timescales
\citep[e.g.,][]{Manmoto_1996a,Lyubarskii_1997a,Nowak_1999b,Kotov_2001a,Arevalo_2006a}.
This class of models can be traced back to the original work of
\citet{Lightman_1974a} and \citet{Shakura_1976a} on disk instabilities
as the source for black hole variability. There are clear indications
that the variability in the \mbox{X-ray} regime observed during the
hard state (for $\Gamma_1 \lesssim 2.65$) has its origin at low
energies.  The time lags between the four energy bands used here show
that the variability in the harder bands lag those of the softer
energy bands.  The variability is thus induced at energies within or
below the 2.1--4.5\,keV band. Extending the timing analysis to lower
energies not accessible with RXTE, \citet{Uttley_2011a} showed that
especially low-frequency variability originates from even lower
energies. These authors find that in several \mbox{X-ray} binaries,
including \mbox{Cyg~X-1}, variability in the 0.125--0.5\,Hz band
originates from energies $\lesssim$0.5\,keV.

As an extension of propagation models,
\citet{Ingram_2011a,Ingram_2012a} and \citet{Ingram_2013a} have
developed a model based on the truncated disk paradigm for the
spectral states. This model incorporates the propagating fluctuation
origin for the broadband shape of the power spectra and Lense-Thirring
precession for narrow QPOs \citep{Psaltis_2000a}. The model does,
however, not yet produce the double-humped shape of the hard-state
PSDs, but the authors point out that this may be because it does not
yet include variability of the disk itself \citep{Wilkinson_2009a} or
magnetohydrodynamical (MHD) effects. The authors do not explicitly
address energy dependence, but the fact that the higher frequency
variability is produced at smaller radii, where the spectrum is
hardest, may explain the observed prominence of component~2 over
component~1 in the harder PSDs in \mbox{Cyg~X-1}
(Fig.~\ref{fig:psd_grid}).  The $f^{-0.7}$-shape of the time lag
spectra can be reproduced, but not the overlaying structure
(Fig.~\ref{fig:lag_grid}).

A family of models based on the influence of the magnetic fields, the
so-called Accretion-Ejection Instability (AEI) has been proposed to
explain the different kinds of low frequency quasi-periodic
oscillations in microquasars
\citep{Tagger_1999a,Varniere_2012a}. Although in the original version
of the AEI the broad band variability is not studied so that an
application to \mbox{Cyg~X-1} is not straightforward, recent attempts
have been made to look at a broader picture, including the behavior of
high frequency QPO, and proposing an explanation of the different
source spectral states based on different flavors of the AEI
\citep{Varniere_2011a}.

In an alternative approach, \citet{Reig_2003a} explain the \mbox{X-ray}
variability by Compton-upscattering in a jet. This model and its
successive refinements can explain the $f^{-0.7}$-shape of the time
lag spectra \citep{Reig_2003a, Giannios_2004a}, but do not explicitly
address the double-humped structure. \citet{Giannios_2004a} reproduce
the hardening of the high frequency part of the PSDs with increasing
energy \citep[][see also Fig.~\ref{fig:psd_grid}]{Nowak_1999a}.
Further extensions by \citet{Kylafis_2008a} reproduce the shape and
normalization of the $\delta t_\mathrm{avg}$-$\Gamma$ correlation of
\citet{Pottschmidt_2003b} in the hard and intermediate states and
qualitatively obtain a correlation between the peak frequency of the
Lorentzian and $\Gamma$. The model can, however, not explain the
existence of multiple Lorentzian components and does not explicitly
address the soft states, although it could explain them assuming that
the jet is present but its radio and \mbox{X-ray} emission suppressed.
\citet{Uttley_2011a} discuss further problems this model faces when
addressing the soft lags at soft energies outside the RXTE range
presented here.

With new computational developments, timing properties can also be
assessed in more fundamental simulations, most notably in the recent
work of \citet{Schnittman_2013a}, who describe accretion onto a
non-rotating black hole with a global radiation transport code coupled
to a general relativistic MHD simulation.
\citeauthor{Schnittman_2013a} simulate an accretion disk that within
the MHD model self-consistently gives rise to a corona. Detailed
comparisons with high resolution observational studies are premature,
however, especially since the simulations do not yet probe the range
of temporal frequencies presented here. It is reassuring, however that
despite these limitations \citet{Schnittman_2013a} already see some
interesting trends. For example, they are able to reproduce
qualitatively the increase of fractional rms with decreasing mass
accretion rate and therefore luminosity. Our observations show that
the behavior of \mbox{Cyg~X-1} is more complicated, but the simulations
represent an important step towards more detailed studies including
those of time lags between different energy bands.

The different \mbox{X-ray} timing properties in the soft and hard state
suggest the intriguing possibility that some variability properties
could be tied to the presence of the jet or some underlying mechanism
that influences both the jet and the generation of the variability. In
their general relativistic magnetohydrodynamic simulations of
magnetically choked accretion flows around black holes,
\citet{McKinney_2012a} study the constraints under which jets are
launched from an accretion flow. They find connections between
the jet and the disk, including a jet-disk QPO with a period of
$\sim$$70GMc^{-3}$. For \mbox{Cyg~X-1}, with $14.8\pm1.0M_\odot$
\citep{Orosz_2011a}, however, this frequency is on the order of
$\sim$200\,Hz and thus not in the range where we see the strong
broadband variability. The model could be of relevance for the QPOs
seen in other sources.

\section{Summary}\label{sect:summary}

In the following we briefly summarize the highlight results of this
paper that we expect to have the strongest constraints on future
theoretical models and to offer the strongest stimuli for future
investigations of the variability of \mbox{X-ray} binaries:
\begin{itemize}
\item The fractional rms shows a complex behavior that strongly
  depends on the spectral state of the source and on the considered
  energy range. In particular, the fractional variability in the soft
  state is low at energies below $\sim$4.5\,keV ($\sim$15\%) but
  increases with energy up to $\sim$40\% in the 9.4--15\,keV
  range. The ratio of fractional rms at different energies shows a
  tight dependence on spectral shape.
\item The shape of the power spectra is highly dependent on the energy
  band used, with a striking change of behavior at $\Gamma_1 \approx
  2.6$--2.7. This reiterates the importance of an energy-dependent
  approach to PSDs and of carefully addressing the scaling of typical
  energies with mass when comparing variability from different source
  types.
\item The power spectra in hard and intermediate states are dominated
  by two main variability components that shift to higher frequencies
  as the source spectrum softens. In the $\Gamma_1 \approx 2.4$--2.7
  range, the 9.4--15\,keV band is dominated by the higher frequency
  component, while in the 2.1--4.5\,keV and the 4.5--5.7\,keV bands
  the lower frequency component is more prominent.
\item The power spectra of the hardest states show further variability
  components at higher Fourier frequencies. In the 0.125--256\,Hz range,
  these PSDs can easily be confused with the softest observations.
\item The coherence is close to unity at frequencies below
  $\sim$10\,Hz in the hard and soft states. At the transition from
  intermediate to soft state, the coherence drops strongly, with hints
  of structure similar to the two dominant components of the power
  spectra. It recovers in the soft state. Above $\sim$10\,Hz, the
  coherence cannot be constrained in our orbit-wise approach.
\item The time lag spectra in the hard and intermediate states show
  features with a double-humped structure similar to the two dominant
  components of the PSDs and with the same evolution to higher
  frequencies as the photon spectra soften. No structure is visible in
  the time lag spectra with soft photon spectra ($\Gamma_1 \gtrsim
  2.7$).
\item The average time lags in the 3.2--10\,Hz band show a non-linear
  increase with $\Gamma_1$ until $\Gamma_1 \approx 2.65$. For
  $\Gamma_1 \gtrsim 2.65$, the average lags are low and show no
  dependence on spectral shape.
\end{itemize}
We have shown that taking into account spectro-timing correlation does
not only provide a more holistic description of source properties, but
can also help to assess the quality of spectral models. The
spectro-timing analysis presented here for the exceptional data set of
\mbox{Cyg~X-1} can be used as a template for other sources with a
worse coverage of different spectral shapes. The behavior it describes
will allow to test future theoretical models for accretion and
ejection processes against observations. Better constraints of some
elusive timing parameters, such as the detailed structure of time lag
spectra in individual observations, require larger area instruments
and further monitoring of the long-term evolution as may be
provided with future satellites such as ASTROSAT \citep{Paul_2013a}
and NICER \citep{Gendreau_2012a}.

\begin{acknowledgements} This work has been partially funded by the
  Bundesministerium f\"ur Wirtschaft und Technologie under Deutsches
  Zentrum f\"ur Luft- und Raumfahrt Grants 50\,OR\,1007 and
  50\,OR\,1113 and by the European Commission through ITN 215212
  ``Black Hole Universe''. It was partially completed by LLNL under
  Contract DE-AC52-07NA27344, and is supported by NASA grants to LLNL
  and NASA/GSFC. Support for this work was also provided by NASA
  through the Smithsonian Astrophysical Observatory (SAO) contract
  SV3-73016 to MIT for Support of the Chandra X-Ray Center (CXC) and
  Science Instruments; CXC is operated by SAO for and on behalf of
  NASA under contract NAS8-03060. We further acknowledge the support
  by the DFG Cluster of Excellence "Origin and Structure of the
  Universe" and are grateful for the support by MCB through the
  Computational Center for Particle and Astrophysics (C2PAP). This
  research has made use of NASA's Astrophysics Data System
  Bibliographic Services. We thank John E. Davis for the development
  of the \texttt{slxfig} module used to prepare all figures in this
  work and Fritz-Walter Schwarm and Ingo Kreykenbohm for their work on
  the Remeis computing cluster. This research has made use of ISIS
  functions (\texttt{isisscripts}) provided by ECAP/Remeis observatory
  and MIT\footnote{\url{http://www.sternwarte.uni-erlangen.de/isis/}}.
  Without the hard work by Evan Smith to schedule the \mbox{Cyg~X-1}
  so uniformly for more than a decade, this paper would not have been
  possible. VG is grateful for the support through the ESAC faculty
  grant program to support student attendance to the workshop
  ``Spectral/timing properties of accreting objects: from \mbox{X-ray}
  binaries to AGN'' that proved pivotal for the basic idea behind this
  research.
\end{acknowledgements}

\bibliographystyle{aa} 
\bibliography{mnemonic,aa_abbrv,references}

\appendix
\section{Overview of the average PSDs at different spectral shapes}

\begin{figure*}
\includegraphics[width=\textwidth]{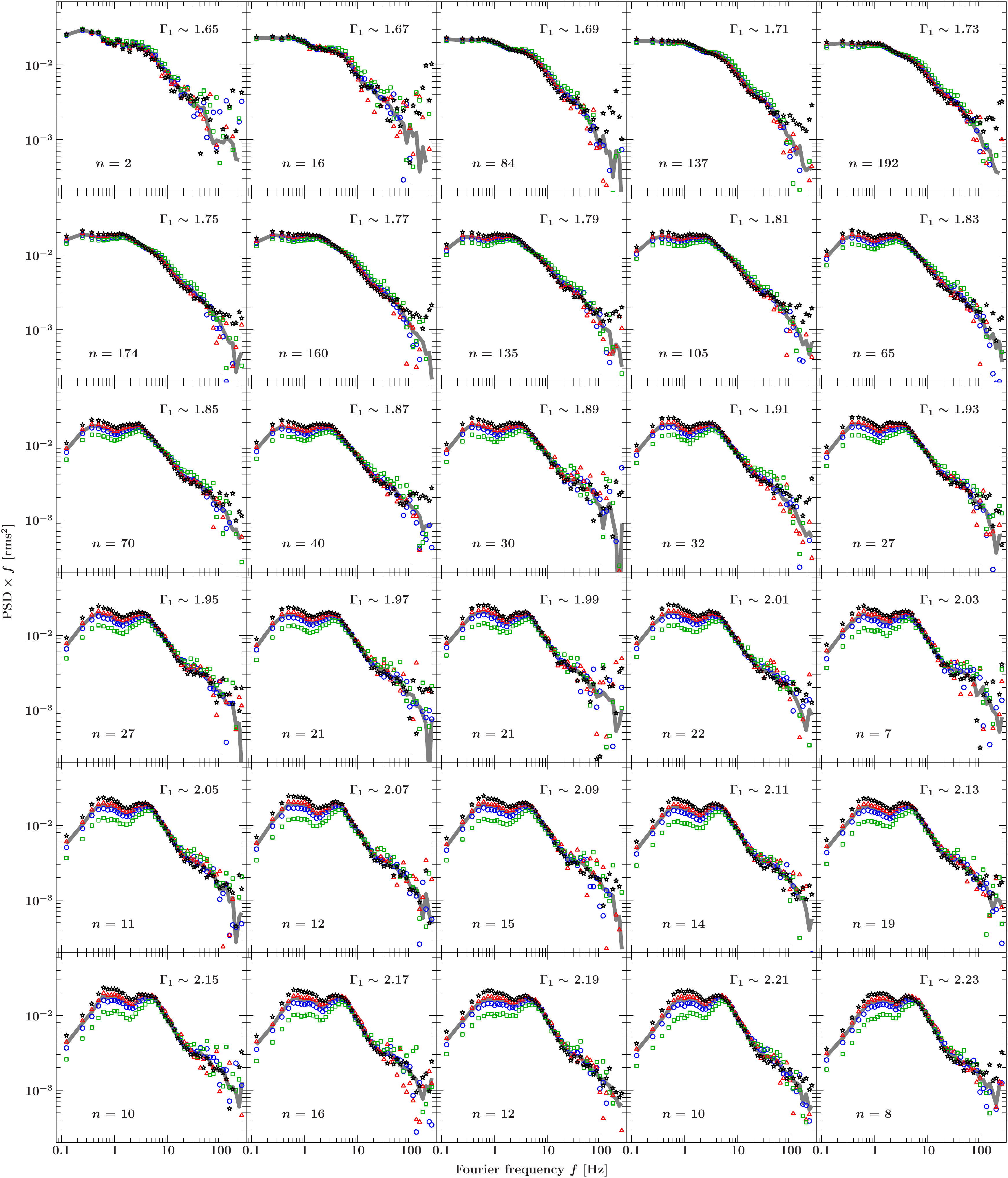}
\caption{Overview of the average PSDs at different spectral shapes.
  PSDs are calculated on a logarithmically binned grid with $\mathrm{d}
  f/f = 0.15$. Each PSD is the average of all $n$ PSDs falling within
  the $\Gamma_1 \pm 0.01$ interval for the given $\Gamma_1$ values.
  Black stars show PSDs in the 2.1--4.5\,keV band, red triangles in
  the 4.5--5.7\,keV band, blue circles in the 5.7--9.4\,keV, and green
  squares in the 9.4--15\,keV band. Gray line represents the PSD in
  the total 2.1--15\,keV band.}\label{fig:avg_psds1}
\end{figure*}

\begin{figure*}
\includegraphics[width=\textwidth]{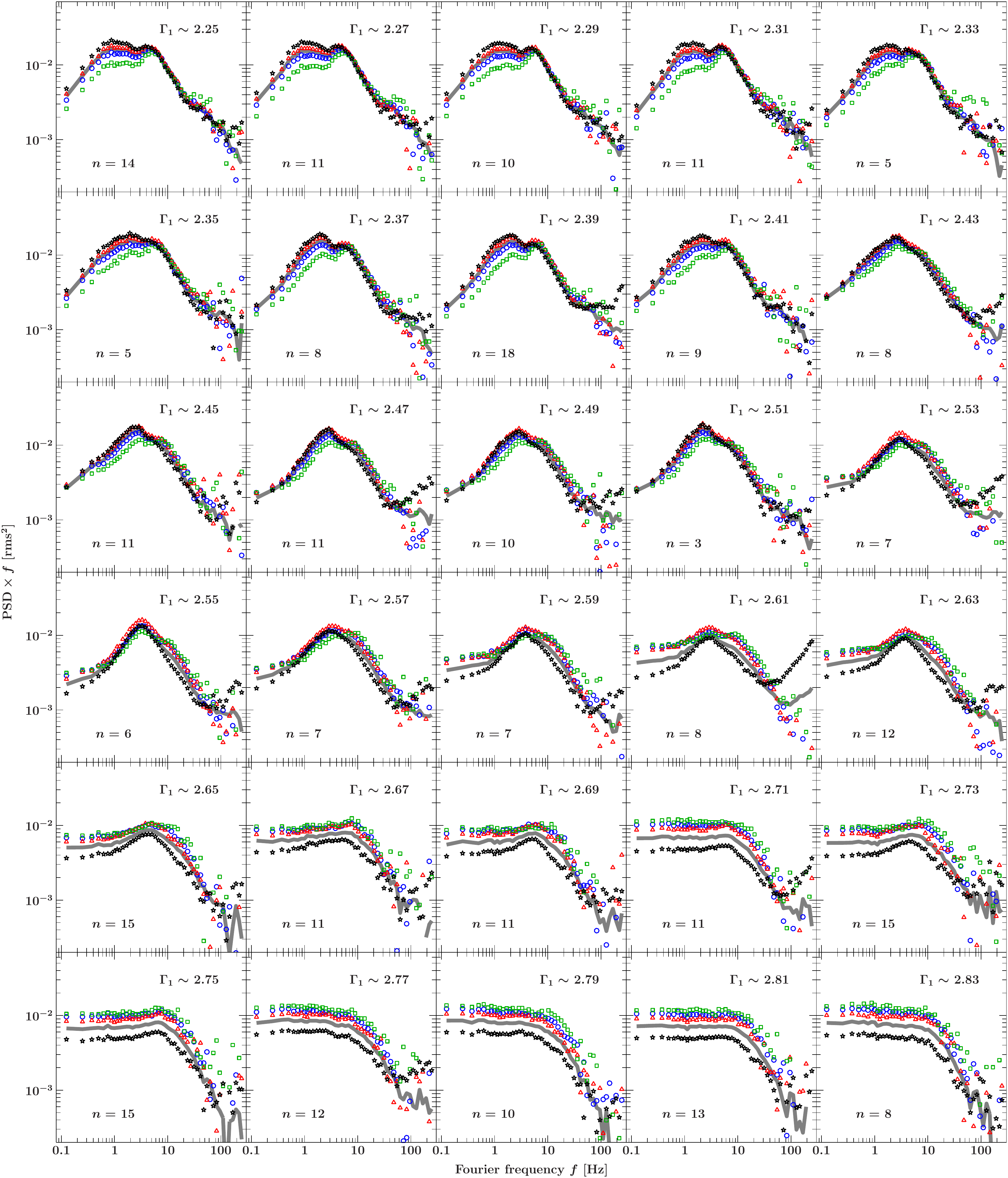}
\caption{Overview of the average PSDs at different spectral shapes.
  PSDs are calculated on a logarithmically binned grid with $\mathrm{d}
  f/f = 0.15$. Each PSD is the average of all $n$ PSDs falling within
  the $\Gamma_1 \pm 0.01$ interval for the given $\Gamma_1$ values.
  Black stars show PSDs in the 2.1--4.5\,keV band, red triangles in
  the 4.5--5.7\,keV band, blue circles in the 5.7--9.4\,keV, and green
  squares in the 9.4--15\,keV band. Gray line represents the PSD in
  the total 2.1--15\,keV band. The increase in the 2.1--4.5\,keV
  PSDs seen at highest frequencies (especially for $\Gamma_1 = 2.61$)
  is not a source intrinsic effect, but due to the high flux in this
  band and resulting telemetry overflow. }\label{fig:avg_psds2}
\end{figure*}

\begin{figure*}
\includegraphics[width=\textwidth]{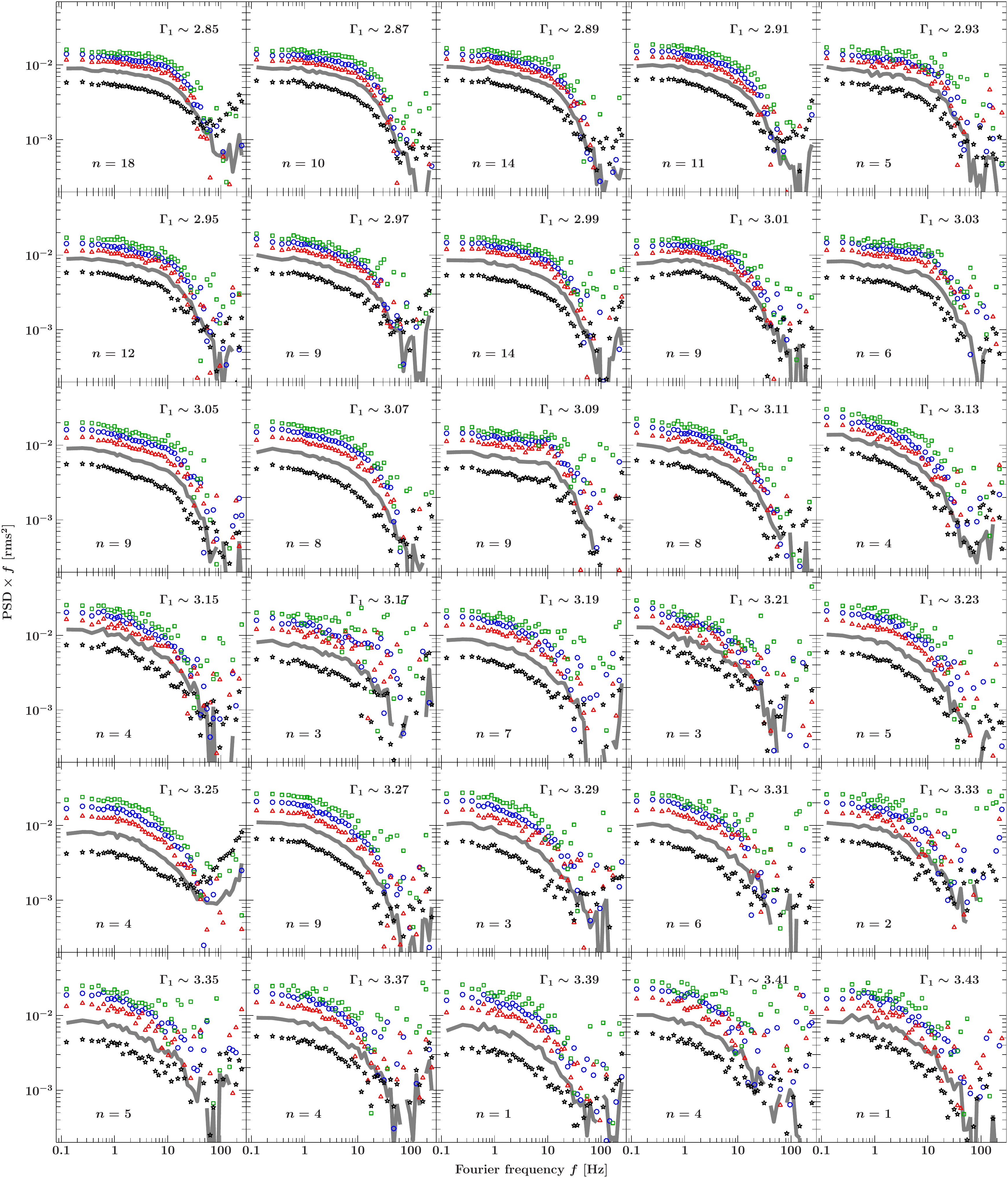}
\caption{Overview of the average PSDs at different spectral shapes.
  PSDs are calculated on a logarithmically binned grid with $\mathrm{d}
  f/f = 0.15$. Each PSD is the average of all $n$ PSDs falling within
  the $\Gamma_1 \pm 0.01$ interval for the given $\Gamma_1$ values.
  Black stars show PSDs in the 2.1--4.5\,keV band, red triangles in
  the 4.5--5.7\,keV band, blue circles in the 5.7--9.4\,keV, and green
  squares in the 9.4--15\,keV band. Gray line represents the PSD in
  the total 2.1--15\,keV band. The increase in the 2.1--4.5\,keV
  PSDs seen at highest frequencies (especially for $\Gamma_1 = 3.25$)
  is not a source intrinsic effect, but due to the high flux in this
  band and resulting telemetry overflow.}\label{fig:avg_psds3}
\end{figure*}

\begin{figure*}
\includegraphics[width=\textwidth]{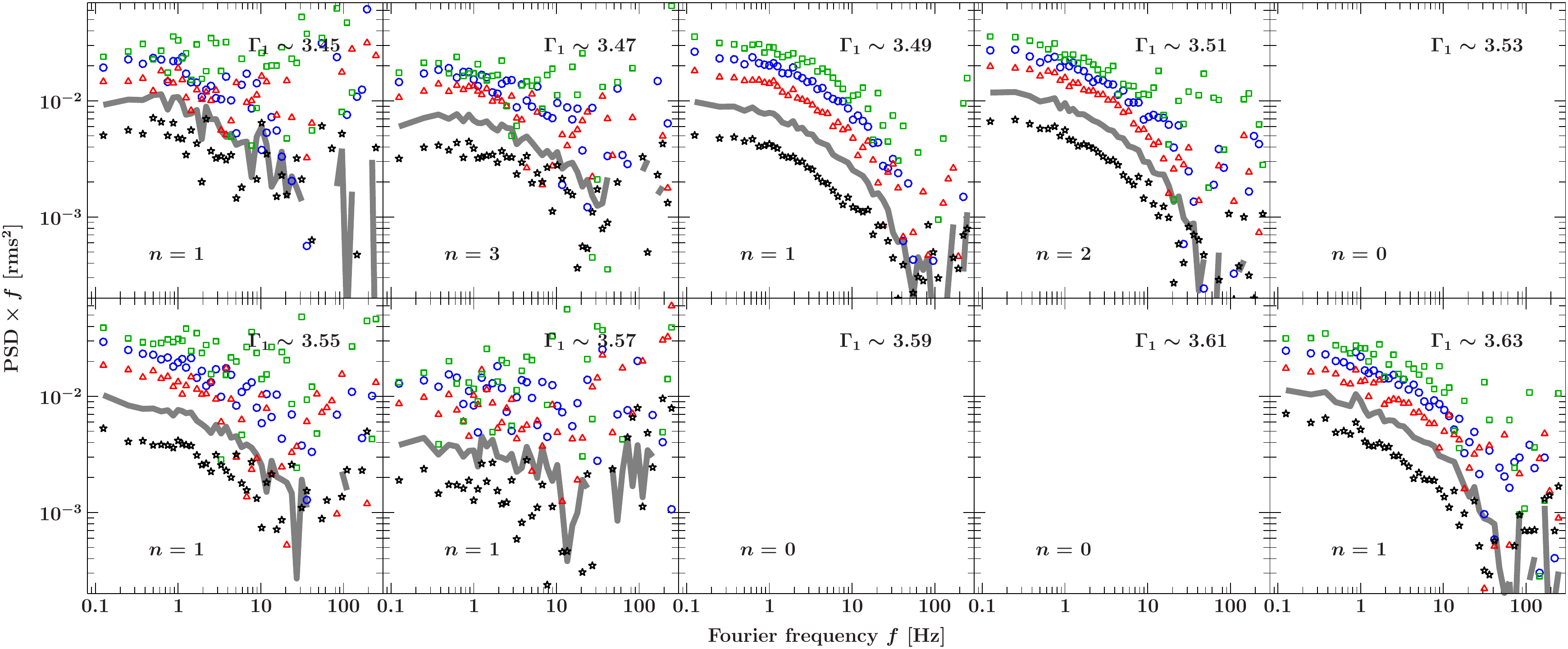}
\caption{Overview of the average PSDs at different spectral shapes.
  PSDs are calculated on a logarithmically binned grid with $\mathrm{d}
  f/f = 0.15$. Each PSD is the average of all $n$ PSDs falling within
  the $\Gamma_1 \pm 0.01$ interval for the given $\Gamma_1$ values. 
  Black stars show PSDs in the 2.1--4.5\,keV band, red triangles in
  the 4.5--5.7\,keV band, blue circles in the 5.7--9.4\,keV, and green
  squares in the 9.4--15\,keV band. Gray line represents the PSD in
  the total 2.1--15\,keV band.}\label{fig:avg_psds4}
\end{figure*}

\end{document}